\documentclass[journal]{IEEEtran}

\usepackage{cite}
\usepackage{amsmath,amssymb,amsfonts,mathrsfs}
\usepackage{algorithmic}
\usepackage{graphicx}
\usepackage{subfigure}
\usepackage{textcomp}

\usepackage{bm}
\usepackage{cases}
\usepackage{hyperref}
\usepackage[ruled,linesnumbered,vlined]{algorithm2e}

\begin{document}

\title{\huge Robust Trajectory and Offloading for Energy-Efficient UAV Edge Computing in Industrial Internet of Things}
\author{Xiao Tang, Hongrui Zhang, Ruonan Zhang, Deyun Zhou, Yan Zhang, and Zhu Han
\thanks{X. Tang, H. Zhang, R. Zhang, and D. Zhou are with the School of Electronics and Information, Northwestern Polytechnical University, Xi'an 710072, China (email: tangxiao@nwpu.edu.cn). Y. Zhang is with the Department of Informatics, University of Oslo, 0316 Oslo, Norway. Z. Han is with the Department of Electrical and Computer Engineering at the University of Houston, Houston, TX 77004 USA.}
}

\maketitle

\begin{abstract}
Efficient data processing and computation are essential for the industrial Internet of things (IIoT) to empower various applications, which yet can be significantly bottlenecked by the limited energy capacity and computation capability of the IIoT nodes. In this paper, we employ an unmanned aerial vehicle (UAV) as an edge server to assist IIoT data processing, while considering the practical issue of UAV jittering. Specifically, we propose a joint design on trajectory and offloading strategies to minimize energy consumption due to local and edge computation, as well as data transmission. We particularly address the UAV jittering that induces Gaussian-distributed uncertainties associated with flying waypoints, resulting in probabilistic-form flying speed and data offloading constraints. We exploit the Bernstein-type inequality to reformulate the constraints in deterministic forms and decompose the energy minimization to solve for trajectory and offloading separately within an alternating optimization framework. The subproblems are then tackled with the successive convex approximation technique. Simulation results show that our proposal strictly guarantees robustness under uncertainties and effectively reduces energy consumption as compared with the baselines.
\end{abstract}


\section{Introduction} \label{sec:1}

The rapid development of the Internet of Things (IoT) technology and network infrastructure empowers the informatization and intelligentization of conventional industrial plants and applications, leading to the industrial IoT (IIoT) for digital transformation~\cite{IIoT}. Towards this vision, various-type and massive-number of radio-frequency identification and sensorial nodes have been deployed, thus enabling the sensing, communication, and computation capability, and establishing the foundation for IIoT. These sensorial nodes are usually with limited energy capacity and computing capability, yet they are expected for long-time operations in certain punishing physical environments. With IIoT getting a wealth of operational and actionable data in industrial environments, they help improve the productivity with maintained quality and increase the revenue with reduced cost~\cite{IIoTnode}.

As IIoT connects various machines and devices in industries, data processing becomes more important, as the failure or delayed data response may probably have more catastrophic consequences as compared with general IoT. However, the IIoT data are usually in denser and more complicated forms, and thus the processing probably goes beyond the capability of the IIoT devices and also expects more energy consumption~\cite{edge}. Challenged by the intensive requirement of data processing in IIoT environments, unmanned aerial vehicle (UAV)-enabled edge computing has been leveraged as an alternative while effective approach. UAVs can be flexibly deployed to the vicinity of the IIoT nodes to assist the data processing. Different processors can be conveniently mounted onto UAVs, providing on-demand computation capability, which is significantly superior to that of the IIoT nodes and satisfying the computation requirements in different IIoT scenarios~\cite{uavedge}.

The higher requirements of data computation also lead to increased energy demands, raising a challenging issue in IIoT contexts. In this respect, thanks to the diverse energy sources and flexible charging for UAVs, UAV-enabled edge computing reduces energy consumption at the capacity-limited and charging-inconvenient IIoT nodes~\cite{energy}. Nevertheless, onboard batteries at both UAVs and IIoT nodes are still limited, and thus energy conservation and efficient utilization are pivotal for self-sustainable and effective operations in IIoT. Moreover, given the fact that industries account for a large portion of carbon footprint, the energy optimization in IIoT helps reduce emission intensities and thus contributes to the evolution towards future greener industries~\cite{green}.

Due to the benefits of UAV-enabled edge computing for IIoT, while facing with energy challenges, there has emerged rich literature addressing the energy-related issues in the respect of mobility design, deployment optimization, access control, resource allocation, and so forth~\cite{uavmec}. Despite their effectiveness, they are mostly with the underlying assumption of perfect physical conditions while comparatively few have considered the uncertainties therein. Particularly, UAV jittering is a fairly common phenomenon in practical applications, which is quite likely to affect the system operation substantially. The jittering may result from various factors including inaccurate positioning information, imperfect flying control, air turbulence, human intervention, etc. Therefore, jittering is inevitable and needs to be tackled properly for performance guarantee. In this respect, the existing works mostly aim for robustness specially addressing the worst case. However, in real practice, the worst case usually occurs with a rather low probability and thus the worst case-oriented approach may, to a certain extent, mislead the strategy for the common cases, which usually occur with a higher probability. Instead, the distributionally robust approach arises as an attractive solution to tackle the uncertainties according to the probability rather than solely accounting for the worst case.

Towards the aforementioned issues, we in this paper intend to minimize the energy consumption for the UAV-assisted computation while considering the UAV jittering. In particular, we jointly optimize the UAV trajectory and offloading strategy to minimize energy consumption concerning local computing, data transmission, and edge computing. The jittering affects the UAV flying as well as offloading, for which robust performance is guaranteed in a probabilistic sense. In particular, the main contributions can be summarized as follows:
\begin{itemize}
	\item We propose an energy-minimized UAV edge computing scheme for IIoT in the presence of UAV jittering. The jittering imposes random geographic derivations on the desired flying waypoints of the UAV, and further affects the UAV trajectory as well as computation offloading.
	\item We model uncertainties associated with UAV flying waypoints with the Gaussian model, and then establish the chance-constrained robust performance guarantees. The constraints in probabilistic forms are then converted into deterministic counterparts to facilitate the analysis by exploiting the Bernstein-type inequality.
	\item The reformulated energy minimization is decoupled into subproblems to tackle trajectory and offloading, respectively. The subproblems are then solved through successive convex approximation (SCA), and the alternating iterations between solving the subproblems provide the strategy to minimize the overall energy consumption.
\end{itemize}

The rest of this paper is organized as follows. Sec.~\ref{sec:2} reviews the related work. Sec.~\ref{sec:3} introduces the model of UAV-assisted data computation offloading for IIoT. Sec.~\ref{sec:4} formulates the energy minimization problem in the presence of UAV jittering along with the algorithm design. Sec.~\ref{sec:5} provides the simulation results, and this paper is concluded in Sec.~\ref{sec:6}.

\section{Related Work} \label{sec:2}

As IIoT empowers and accelerates the digitalization of conventional industrial facilities, efficient data processing is the essential ingredient and attracts wide research interests~\cite{iiotdata}. In this context, the UAV edge computing, featuring on-demand deployment and flexible configuration, has prevailed as an efficient solution to remedy the resource shortage in IIoT~\cite{uavmec}. In~\cite{uavmec1}, the authors propose a UAV edge system serving multiple IIoTs, with a learning-aided resource allocation strategy to minimize the response time on data monitoring. In~\cite{uavmec2}, the authors propose a UAV-aided ultra-reliable and low-latency computation offloading for mission-critical IoT applications, where the UAV locations and offloading decisions are jointly optimized. In~\cite{uavmec3}, the authors consider the data collection and computation issues in the UAV-aided IoT system, intending to maintain the data freshness with queuing-based analysis. In~\cite{r1add}, the authors propose to minimize the maximum delay during UAV-enabled edge computing, by jointly investigating the UAV trajectory, offloading task ratio and user scheduling. In~\cite{uavmec4}, the authors investigate the UAV-aided data processing in industrial emergencies, with joint consideration of data computation and storage. In~\cite{r4add2}, the authors propose the reinforcement learning-based approaches for secured computation offloading to the UAV edge. In~\cite{r4add1}, the authors maximize the secure computation efficiency in UAV edge system by jointly optimizing offloading decision and resource management.

Meanwhile, as the fundamental to power UAVs and IIoT nodes, energy-related issues also attract considerable research attention~\cite{cja}. In~\cite{energy1}, the authors consider the computation offloading from IIoT to fog servers, while minimizing the energy consumption subject to delay constraints. In~\cite{energy2}, the authors investigate the association and offloading strategy among multiple edge servers and IIoT devices for energy minimization. In~\cite{energy3}, the authors propose an energy-efficient dispersed computing paradigm to jointly optimize the computing tasks and network resources. In~\cite{energy4}, the authors exploit UAVs as edge servers to offload the computation, while jointly optimizing the UAV trajectory, transmission, and computation for the highest energy efficiency. In~\cite{energy5}, the authors investigate the UAV edge computing to minimize the energy consumption and completion time for data computation, while satisfying the task requirements at each ground offloading users. In~\cite{energy6}, the authors intend to minimize the energy consumption of the UAV-IoT network, a reinforcement learning approach is proposed to optimize the trajectory and resources. In~\cite{energy7}, the authors minimize the energy consumption in UAV-aided edge computing system through rate splitting for optimized computing task allocation. In~\cite{aeadd}, the authors exploit the real-time user density data provided by Tencent, China, and determine the 3D location for the UAV edge servers to assist the computation.

The aforementioned studies share the underlying assumption of perfect environmental conditions, whereas the uncertainty issues are relatively less addressed in the existing literature. In~\cite{robust1}, the authors investigate a jittering UAV in air-ground communications and achieve energy-efficient transmissions in the worst case. In~\cite{robust2}, the authors consider the UAV jittering issue and tackle the beam alignment through compressed sensing. In~\cite{robust3}, the authors address the UAV-assisted edge caching and minimize the delay performance within a robust game framework. Besides the robust designs against the uncertainties in the worst case, a recently emerged branch is the robust design that takes into account the distribution of the uncertainties. In~\cite{dro1}, the authors consider the computation offloading in the space-air-ground network, where the uncertainties in arrived task is addressed for latency minimization under energy constraints. In~\cite{dro3}, the authors investigate robust edge intelligence by combining cloud knowledge and local learning. In~\cite{dro2}, the authors propose a distributionally robust offloading strategy for IoT while considering the channel uncertainties. In summary, the UAV edge computing for IIoT under uncertainties is of paramount importance and thus desires thorough investigation, while the existing work mostly focuses on the uncertainties associated with transmission or the worst-case robustness. In contrast, we in this work specially address the UAV jittering-induced uncertainties with explicit consideration of the distribution. The further random consequences on flying and offloading are tackled with a chance-constrained performance guarantee. Thus, the proposed robust approach has the potential to better facilitate practical IIoT applications.

\section{System Model} \label{sec:3}

\begin{figure}[t]
  \centering
  \includegraphics[width=6.5cm]{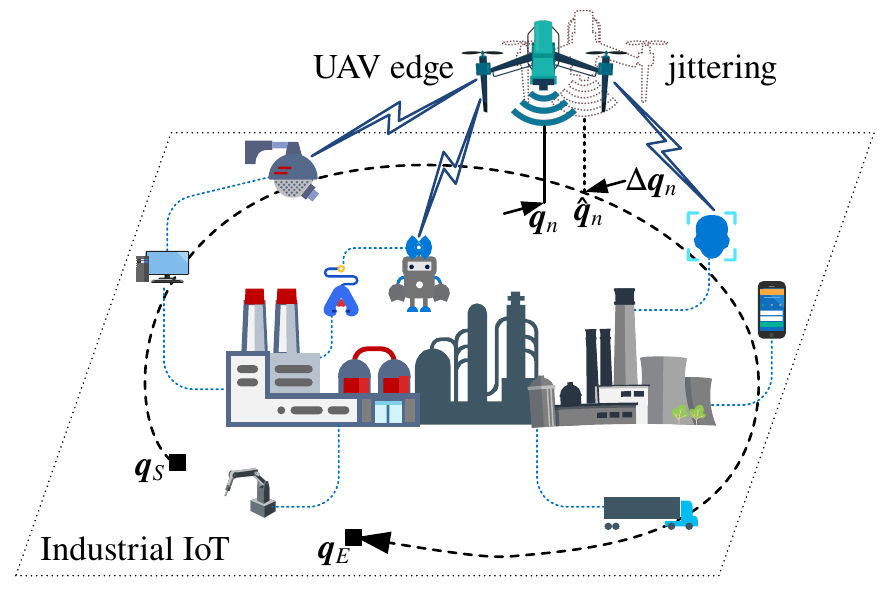}
  \caption{System model.}
  \label{fig:sys}
\end{figure}

We consider an IIoT that consists $ K $ ground nodes, denoted by $ \mathcal{K} = \left\{1,2,\cdots, K\right\} $, located within an area denoted by $\mathcal{A}$, as shown in Fig.~\ref{fig:sys}. The location of IIoT node-$k\in\mathcal{K}$ is denoted by $ \bm{w}_k = \left[ w_k^{{(x)}}, w_k^{{(y)}}, 0 \right] $. The nodes deployed in the IIoT context intend for certain missions, e.g., video stream screening to report delinquency or sensing data processing for event inference. Due to the limited energy storage and computation capability of these nodes, a UAV is dispatched as a mobile edge server to assist the data processing. Specifically, the UAV needs to visit IIoT nodes within a time period given by $ T $, which is equally divided into $ N $ intervals, denoted by $ \mathcal{N} = \left\{1,2,\cdots, N\right\} $, each is of a time length $ T/N $. Correspondingly, the time instance is denoted by $ \bar{\mathcal{N}} = \left\{0,\mathcal{N}\right\} $, where the waypoints of the UAV is denoted by $ \left\{\bm{q}_n\right\}_{n\in\bar{\mathcal{N}}} $. The waypoint $ \bm{q}_n $ is specified as $ \bm{q}_n = \left[ q_n^{(x)}, q_n^{(y)}, H \right] $, where $ H $ is the constant altitude. Also, we assume that $ \bm{q}_0 $ and $ \bm{q}_{N} $ are fixed as
\begin{equation} \label{eq:startend}
	\bm{q}_0 = \bm{q}_S,\quad \bm{q}_{N} = \bm{q}_E,
\end{equation}
where $ \bm{q}_S $ and $ \bm{q}_{E} $ are the starting point and end point, respectively, in accordance with the requirement of edge mission or location of charging stations in real practice.
Moreover, any two consecutive waypoints are subject to the UAV flying constraint specified as
\begin{equation} \label{eq:speedcon}
	\left\| \bm{q}_{n} - \bm{q}_{n-1} \right\| \le V^{\max}\frac{T}{N}, \quad \forall n\in\mathcal{N},
\end{equation}
where $ V^{\max} $ is the maximum flying speed.

\subsection{Offloading Model}
The nodes in the IIoT environment have a certain amount of data requiring further processing, which can be done locally at the nodes or forwarded to the UAV as the flying edge server. In particular, denote $ d_{n,k}^{\text{(loc)}} $ and $ d_{n,k}^{\text{(off)}} $ as the data amount processed at the nodes and offloaded to the UAV, respectively, then at time interval-$n$, the following computation constraint is required
\begin{equation} \label{eq:datacon}
	\sum_{n\in\mathcal{N}} d_{n,k}^{\text{(loc)}} + d_{n,k}^{\text{(off)}} \ge D_k,\quad \forall k\in\mathcal{K},
\end{equation}
where $ D_k $ is the computation data amount at node-$k$. This constraint guarantees that the generated data can be all processed, while the computation tasks can be executed locally or at the edge. Further, for the locally processed data, denote the computation capability of node-$k$ as $ c_k $ in cycle per bit with the maximum computation frequency denoted by $ F_k^{\max} $, and then the local data processing at node-$k$ is subject to the following computation constraint
\begin{equation} \label{eq:loccon}
	c_k d_{n,k}^{\text{(loc)}} \le \frac{T}{N}F_k^{\max}, \quad\forall n\in\mathcal{N},
\end{equation}
which requires the local computation to be finished during the considered time. Meanwhile, for the computation offloading to the UAV, the data first needs to be forwarded to the UAV. Assume line-of-sight communication between the UAV and IIoT nodes, we adopt the free-space propagation model and the channel between the UAV and node-$k$ at time interval-$n$ is given as
\begin{equation} \label{eq:channel}
	h_{n,k} = \frac{\beta_0}{\left\| \bm{q}_n - \bm{w}_k \right\|^2},
\end{equation}
where $ \beta_0 $ is the transmission attenuation at the reference distance of unit meter. Here, we use the location of UAV at time instance-$n$ to approximate that in the whole time interval. This channel model has been widely used in the UAV edge computing scenario~\cite{energy,Tao}, while can be extended to the more generalized segmented ray-tracing channel model as that in~\cite{ray} when the radio map regarding the investigated IIoT scenario is available.
Then, the offloading rate at node-$k$ at time interval-$n$ is
\begin{equation} \label{eq:rate}
	R_{n,k} = B \log \left(1 + \frac{p_{n,k} h_{n,k} }{\sigma_0^2} \right),
\end{equation}
where $ B $ is the uplink bandwidth, $ \sigma_0^2 $ is the background noise power, and the transmit power from node-$k$ in time interval-$n$ is constrained as
\begin{equation} \label{eq:powercon}
    0 \le p_{n,k} \le p^{\max}_k, \quad\forall n\in\mathcal{N},
\end{equation}
with $ p_k^{\max} $ being the maximum allowed power at node-$k$.

We consider the data offloading from different nodes conducted in a time-orthogonal manner, i.e., a portion of $ \tau_{n,k} $ is allocated to node-$k$ at time interval-$n$ satisfying
\begin{equation} \label{eq:taucon} \left\{
\begin{aligned}
	&0 \le \tau_{n,k} \le 1, \quad\forall k\in\mathcal{K}, \quad\forall n\in\mathcal{N}, \\
	&\sum_{k\in\mathcal{K}} \tau_{n,k} \le 1,\quad\forall n\in\mathcal{N}.
\end{aligned} \right.
\end{equation}
The time-orthogonal transmissions can be conveniently implemented as the UAV as the edge server actively determines the scheduling. Also, the UAV can fly near to the scheduled node with guaranteed coverage to facilitate the computation offloading. Accordingly, the offloaded data for edge computing at node-$k$ in time interval-$n$ is subject to the transmission constraint as
\begin{equation} \label{eq:offcon}
	d_{n,k}^{\text{(off)}} \le \tau_{n,k}\frac{T}{N}R_{n,k}, \quad\forall k\in\mathcal{K}, \quad\forall n\in\mathcal{N}.
\end{equation}
For the UAV to process the offloaded data, suppose the computing frequency of the UAV at time interval-$n$ is $ f_n^{\text{(edg)}} $, then UAV edge computing is subject to the computing capability that
\begin{equation} \label{eq:freqcon}
	0 \le f_n^{\text{(edg)}} \le F_0^{\max},\quad n\in\mathcal{N},
\end{equation}
where $F_0^{\max}$ is the highest computing frequency of the UAV. Due to the causality of offloading and computation, we have the following constraint
\begin{equation} \label{eq:edgcon1}
	\sum\limits_{m=2}^{n} \frac{T}{N}f_{m}^{\text{(edg)}} \le c_0 \sum\limits_{m=1}^{n-1}\sum\limits_{k\in\mathcal{K}} d_{m,k}^{\text{(off)}} ,\quad \forall n\in\mathcal{N}\backslash\left\{N\right\},
\end{equation}
which indicates that the actually conducted edge computation until current time interval is no more than the required amount due to the offloaded data until previous interval, with $ c_0 $ being the computation capability of the UAV in cycle per bit. Also, at the last time interval, all the offloaded computation should be finished as
\begin{equation} \label{eq:edgcon2}
	\sum\limits_{m=2}^{N} \frac{T}{N}f_{m}^{\text{(edg)}} \ge c_0 \sum\limits_{m=1}^{N-1}\sum\limits_{k\in\mathcal{K}} d_{m,k}^{\text{(off)}}.
\end{equation}
The constraints in~(\ref{eq:edgcon1}) and~(\ref{eq:edgcon2}) above correspond to the causality and completion of the computation offloading process. In this regard, we actually allow a relatively flexible space for the UAV to plan its computation tasks during the flight, facilitating energy-efficient operations at the UAV edge.

\subsection{Energy Model}
The energy consumption during IIoT data processing now concerns two parties, i.e., the IIoT nodes and the UAV. For the nodes, they divide the data into the local-processing part and offloading part, which incurs the computation energy and transmission energy, respectively. Accordingly, we adopt the dynamic frequency scaling technique based computation energy model~\cite{Tao,dfs}, and the energy for node-$k$ at time interval-$n$ is specified as
\begin{equation} \label{eq:energynode}
	E_{n,k}^{\text{(nod)}} = \frac{\lambda_k \left( c_k d_{n,k}^{\text{(loc)}} \right)^3}{\left(T/N\right)^2} + \tau_{n,k}\frac{T}{N}p_{n,k},
\end{equation}
where $ \lambda_k $ is the effective capacitance coefficient of node-$k$. Similarly, the UAV conducts edge computing with frequency $ f_{n}^{\text{(edg)}} $ during time interval-$n$, where the computation frequency satisfying the inequality in~(\ref{eq:edgcon1}) and~(\ref{eq:edgcon2}) guarantees the causality and completion of the computation offloading. Then, the consumed energy at the UAV edge at time interval-$n$ is
\begin{equation} \label{eq:energyedge}
	E_n^{\text{(edg)}} = \lambda_0 \frac{T}{N} \left(f_n^{\text{(edg)}}\right)^3,
\end{equation}
where $ \lambda_0 $ is the effective capacitance coefficient of the edge server. Note here the computation energy in~(\ref{eq:energynode}) and~(\ref{eq:energyedge}) has different forms as the former is calculated based on computation data amount while the latter is based on computation frequency. 
Therefore, the overall energy consumption for the computation in UAV-aided IIoT context can be obtained as
\begin{equation} \label{eq:energy}
	E = \sum\limits_{n\in\mathcal{N}}\sum\limits_{k\in\mathcal{K}}E_{n,k}^{\text{(nod)}} + \chi \sum\limits_{n\in\mathcal{N}}E_n^{\text{(edg)}},
\end{equation}
where $ \chi $ is the weight to balance the energy at different parties since the energy sources can be rather different in the aspects of availability, price, recharging, and so on. In our formulation, the weight is set as $ 0 < \chi < 1 $, i.e., smaller than one, so as to better motivate the computation offloading to the UAV edge. Also note that the UAV propulsion energy is not explicitly considered as it can be rather high and generally overshadows the computation energy consumption, while we in this work concentrate on the computation-related strategy design.

\subsection{Uncertainty Model}

Furthermore, we consider the UAV jittering and its influence on the computing offloading. Particularly, in practical scenarios, there exists multi-fold randomness affecting the UAV flying, such as inaccurate localization, air turbulence, imperfect flying control, etc. The randomness induces unpredictable derivations from the desired flying waypoints and the resultant uncertainties for flying and data offloading need to be addressed towards a robust edge computing system. Correspondingly, the flying waypoints of UAV are modeled with uncertainties as
\begin{equation} \label{eq:qdelta}
	\bm{q}_n = \hat{\bm{q}}_n + \Delta\bm{q}_n,\quad \forall n\in\mathcal{N},
\end{equation}
where $ \hat{\bm{q}}_n $ is the desired waypoints and $\Delta\bm{q}_n$ is the associated uncertainty. Due to the numerous random events from various contributing factors, the uncertainty is modeled as a random variable with Gaussian distribution, given as
\begin{equation} \label{eq:qdist}
	\Delta\bm{q}_n \sim \mathscr{N}\left(0, \epsilon_0^2 \bm{I}\right)\quad \forall n\in\mathcal{N},
\end{equation}
where $ \bm{I} $ is a three-by-three identity matrix corresponding to the three dimensions in space. Note that although we have assumed fixed UAV altitude in the system model, there are still uncertainties in the vertical dimension and thus the actual altitude of the UAV is also random. In accordance with the jittering UAV with waypoint uncertainties, the aforementioned constraints concerning the waypoints need to be modified. In particular, the UAV flying constraint is reformulated as
\begin{equation} \label{eq:speedprob}
    \mathbb{P}_{\Delta\bm{q}_n}\left\{ \left\| \bm{q}_{n} - \bm{q}_{n-1} \right\| \le V^{\max}\frac{T}{N} \right\} \ge 1 - \rho_{n}^{\text{(trj)}} , \quad \forall n\in\mathcal{N},
\end{equation}
where $ \rho_{n}^{\text{(trj)}} $ is the probability that the UAV flying violates the maximum speed constraint in time interval-$n$ and the probability is derived based on the uncertainty. Similarly, as the transmission rate from the nodes to the UAV depends on the flying waypoints, the offloading constraint also requires reformulation, given as
\begin{equation} \label{eq:offprob}
    \mathbb{P}_{\Delta\bm{q}_n}\left\{ d_{n,k}^{\text{(off)}} \le \tau_{n,k}\frac{T}{N}R_{n,k} \right\} \ge 1 - \rho_{n,k}^{\text{(off)}}, \quad\forall k\in\mathcal{K}, \:\:\forall n\in\mathcal{N}, 
\end{equation}
where $ \rho_{n,k}^{\text{(off)}} $ is probability that the actual transmission rate fails the requirement of offloading at node-$k$ in time interval-$n$.

\section{Robust Optimization Against UAV Jittering} \label{sec:4}

\subsection{Problem Formulation and Decomposition}

For the considered IIoT system with UAV edge computing, we intend to minimize the energy consumption to prolong the network lifetime. Accordingly, we employ a UAV as a mobile edge server to assist with the offloading tasks. Therefore, the problem is formulated to minimize the system energy consumption while considering the constraints on transmission, scheduling, offloading, and flying with uncertainties, given as
\begin{IEEEeqnarray}{cl}
  \IEEEyesnumber\label{eq:problem} \IEEEnosubnumber*
  \min_{\substack{\left\{ \hat{\bm{q}}_n, f_n^{\text{(edg)}} \right\}_{n\in\mathcal{N}}\\ \left\{d_{n,k}^{\text{(loc)}}, d_{n,k}^{\text{(off)}}\right\}_{k\in\mathcal{K},n\in\mathcal{N}}\\ \left\{p_{n,k}, \tau_{n,k}\right\}_{k\in\mathcal{K},n\in\mathcal{N}} }} \quad & E  \\
   \rm{s.t.} \quad & (\ref{eq:startend}), (\ref{eq:datacon}),(\ref{eq:loccon}),(\ref{eq:powercon}),(\ref{eq:taucon}), \nonumber \\
   & (\ref{eq:freqcon}),(\ref{eq:edgcon1}),(\ref{eq:edgcon2}),(\ref{eq:speedprob}),(\ref{eq:offprob}). \nonumber
\end{IEEEeqnarray}
In the formulated problem, we have adopted the probabilistic constraint due to the UAV jittering on speed and offloading as given in~(\ref{eq:speedprob}) and~(\ref{eq:offprob}), rather than the original deterministic counterparts as in~(\ref{eq:speedcon}) and~(\ref{eq:offcon}).

The formulated problem is rather complicated with three-fold difficulties. First, there are multiple contributing factors, including the UAV flying, transmission, scheduling, and offloading, coupled with each other to jointly affect the system energy consumption. Second, the uncertainties associated with UAV flying waypoints are inherently built within the channel model and further result in complicated influence on the offloading operations. Third, the chance-constrained flying and offloading in probabilistic forms hinder efficient treatment in conventional manners for the energy-minimization problem.

To solve the problem effectively, we need to first tackle the constraints in probabilistic forms. In this respect, we exploit the Bernstein-type inequality to convert the constraints into deterministic forms to facilitate the discussion~\cite{b}. Specifically, for the UAV flying constraint~(\ref{eq:speedprob}), by substituting the waypoint uncertainties in~(\ref{eq:qdelta}) into the speed condition, i.e., the term within the probability in~(\ref{eq:speedprob}), we have that
\begin{equation} \label{eq:speeduncer1}
\begin{aligned}
   &\left\| \Delta\bm{q}_n - \Delta\bm{q}_{n-1} \right\|^2 + 2\left( \hat{\bm{q}}_n - \hat{\bm{q}}_{n-1} \right)^T \left(\Delta\bm{q}_n - \Delta\bm{q}_{n-1}\right) \\ +& \left\| \hat{\bm{q}}_n - \hat{\bm{q}}_{n-1} \right\|^2 - \left( V^{\max}\frac{T}{N} \right)^2 \le 0, \quad \forall n\in\mathcal{N}.
\end{aligned}
\end{equation}
Then, given the uncertain distribution in~(\ref{eq:qdist}), we can readily derive that $ \left( \Delta\bm{q}_n - \Delta\bm{q}_{n-1} \right) \sim \mathscr{N}\left(0, 2\epsilon_0^2 \bm{I}\right) $, and thus $ \left( \Delta\bm{q}_n - \Delta\bm{q}_{n-1} \right)/\sqrt{2}\epsilon_0 $ follows standard Gaussian distribution and the inequality in~(\ref{eq:speeduncer1}) can be rewritten as
\begin{equation} \label{eq:speeduncer2}
\begin{aligned}
   &-2\epsilon_0^2\left\| \frac{\Delta\bm{q}_n - \Delta\bm{q}_{n-1}}{\sqrt{2}\epsilon_0} \right\|^2 - \left\| \hat{\bm{q}}_n - \hat{\bm{q}}_{n-1} \right\|^2 + \left( V^{\max}\frac{T}{N} \right)^2 \\ &- 2\sqrt{2}\epsilon_0\left( \hat{\bm{q}}_n - \hat{\bm{q}}_{n-1} \right)^T \left( \frac{\Delta\bm{q}_n - \Delta\bm{q}_{n-1}}{\sqrt{2}\epsilon_0}\right)  \ge 0, \quad \forall n\in\mathcal{N}.
\end{aligned}
\end{equation}
The inequality above then satisfies the standard condition of Bernstein-type inequality, which allows converting the original probabilistic condition to the following deterministic form as
\begin{subnumcases}{\label{eq:grouptrj}}
    \begin{aligned}
    &\mathsf{Tr}\left(\bm{A}_n^{(\text{trj})}\right) - \sqrt{-2\ln\left(\rho_n^{(\text{trj})}\right)}\mu_n^{(\text{trj})} \\
    & \quad + \ln\left(\rho_n^{(\text{trj})}\right)\nu_n^{(\text{trj})} + c_n^{(\text{trj})} \ge 0, \quad \forall n\in\mathcal{N}, 
    \end{aligned} \\
    \left\| \begin{array}{c} \mathsf{vec} \left(\bm{A}_n^{(\text{trj})}\right) \\ \sqrt{2}\bm{b}_n^{(\text{trj})} \end{array} \right\| \le \mu_n^{(\text{trj})}, \quad  \forall n\in\mathcal{N}, \\
    \nu_n^{(\text{trj})} \bm{I} + \bm{A}_n^{(\text{trj})} \succcurlyeq 0, \quad \nu_n^{(\text{trj})} \ge 0, \quad  \forall n\in\mathcal{N},
\end{subnumcases}
where
\begin{subnumcases}{}
    \bm{A}_n^{(\text{trj})} = -2\epsilon_0^2 \bm{I}, \\
    \bm{b}_n^{(\text{trj})} = -\sqrt{2}\epsilon_0\left( \hat{\bm{q}}_n - \hat{\bm{q}}_{n-1} \right), \\
    c_n^{(\text{trj})} = - \left\| \hat{\bm{q}}_n - \hat{\bm{q}}_{n-1} \right\|^2 + \left( V^{\max}\frac{T}{N} \right)^2,
\end{subnumcases}
and $ \left\{ \mu_n^{(\text{trj})}, \nu_n^{(\text{trj})}\right\}_{n\in\mathcal{N}} $ are the newly introduced variables towards the deterministic inequalities. It can be easily verified that the deterministic constraints in~(\ref{eq:grouptrj}) are jointly convex with respect to $ \left\{\hat{\bm{q}}_n\right\}_{n\in\mathcal{N}} $, $ \left\{ \mu_n^{(\text{trj})}\right\}_{n\in\mathcal{N}} $, and $ \left\{ \nu_n^{(\text{trj})}\right\}_{n\in\mathcal{N}} $. Moreover, the uploading constraint in the probabilistic form in~(\ref{eq:offprob}) can be tackled in a similar manner. The inequality within the probability in~(\ref{eq:offprob}) guaranteeing the offloading data amount can be reformulated as
\begin{equation}
    \left\| \bm{q}_n - \bm{w}_k \right\|^2 \le \frac{\beta_0}{\left(2^{\frac{d_{n,k}^{\text{(off)}}}{\tau_{n,k}}\frac{N}{TB}-1}\right)\frac{\sigma_0^2}{p_{n,k}}}, \quad\forall k\in\mathcal{K}, \:\forall n\in\mathcal{N},
\end{equation}
by substituting the uploading transmission rate in~(\ref{eq:rate}) into the condition.
With further consideration of the uncertainties on the waypoints, the inequality above is derived as
\begin{equation}
\begin{aligned}
    &\left\| \Delta\bm{q}_n \right\|^2 + 2 \Delta\bm{q}_n^T\left( \hat{\bm{q}}_n - \bm{w}_k \right) + \left\| \hat{\bm{q}}_n - \bm{w}_k \right\|^2 \\ -& \frac{\beta_0}{\left(2^{\frac{d_{n,k}^{\text{(off)}}}{\tau_{n,k}}\frac{N}{TB}-1}\right)\frac{\sigma_0^2}{p_{n,k}}} \le 0, \quad\forall k\in\mathcal{K}, \:\forall n\in\mathcal{N}.
\end{aligned} 
\end{equation}
Then, by tackling the Gaussian-distributed uncertainties with Bernstein-type inequality, we arrive at the following deterministic equivalence as
\begin{subnumcases}{\label{eq:groupoff}}
    \begin{aligned} \label{eq:groupoffa}
    &\mathsf{Tr}\left(\bm{A}_{n,k}^{(\text{off})}\right) - \sqrt{-2\ln\left(\rho_{n,k}^{(\text{off})}\right)}\mu_{n,k}^{(\text{off})} \\
    &  + \ln\left(\rho_{n,k}^{(\text{off})}\right)\nu_{n,k}^{(\text{off})} + c_{n,k}^{(\text{off})} \ge 0, \:\:\forall k\in\mathcal{K}, \:\forall n\in\mathcal{N}, 
    \end{aligned} \\
    \left\| \begin{array}{c} \mathsf{vec} \left(\bm{A}_{n,k}^{(\text{off})}\right) \\ \sqrt{2}\bm{b}_{n,k}^{(\text{off})} \end{array} \right\| \le \mu_{n,k}^{(\text{off})}, \quad\forall k\in\mathcal{K}, \:\forall n\in\mathcal{N}, \label{eq:groupoffb}\\
    \nu_{n,k}^{(\text{off})} \bm{I} + \bm{A}_{n,k}^{(\text{off})} \succcurlyeq 0, \:\: \nu_{n,k}^{(\text{off})} \ge 0, \:\:\forall k\in\mathcal{K}, \:\forall n\in\mathcal{N}, \label{eq:groupoffc}
\end{subnumcases}
where
\begin{subnumcases}{}
    \bm{A}_{n,k}^{(\text{off})} = -\epsilon_0^2 \bm{I}, \\
    \bm{b}_{n,k}^{(\text{off})} = -\epsilon_0\left( \hat{\bm{q}}_n - \hat{\bm{w}}_{k} \right), \\
    c_{n,k}^{(\text{off})} = - \left\| \hat{\bm{q}}_n - \hat{\bm{w}}_{k} \right\|^2 + \frac{\beta_0}{\left(2^{\frac{d_{n,k}^{\text{(off)}}}{\tau_{n,k}}\frac{N}{TB}-1}\right)\frac{\sigma_0^2}{p_{n,k}}},
\end{subnumcases}
and $ \left\{ \mu_{n,k}^{(\text{off})}, \nu_{n,k}^{(\text{off})}\right\}_{k\in\mathcal{K},n\in\mathcal{N}} $ are the newly introduced variables. Also, the deterministic constraints in~(\ref{eq:groupoff}) are jointly convex with respect to $ \left\{\hat{\bm{q}}_n\right\}_{n\in\mathcal{N}} $ and $ \left\{ \mu_{n,k}^{(\text{off})}, \nu_{n,k}^{(\text{off})}\right\}_{k\in\mathcal{K},n\in\mathcal{N}} $.

With the above treatment regarding the probabilistic constraints due to UAV jittering, we can then rewrite the energy minimization problem in~(\ref{eq:problem}) as
\begin{IEEEeqnarray}{cl}
  \IEEEyesnumber\label{eq:problemeqv} \IEEEnosubnumber*
  \min_{\substack{\left\{ \hat{\bm{q}}_n, f_n^{\text{(edg)}} \right\}_{n\in\mathcal{N}}, \left\{p_{n,k}, \tau_{n,k}\right\}_{k\in\mathcal{K},n\in\mathcal{N}}\\  \left\{d_{n,k}^{\text{(loc)}}, d_{n,k}^{\text{(off)}}\right\}_{k\in\mathcal{K},n\in\mathcal{N}} \\ \left\{ \mu_n^{(\text{trj})}, \nu_n^{(\text{trj})}\right\}_{n\in\mathcal{N}}, \left\{ \mu_{n,k}^{(\text{off})}, \nu_{n,k}^{(\text{off})}\right\}_{k\in\mathcal{K},n\in\mathcal{N}} }} \quad & E  \\
   \rm{s.t.} \quad & (\ref{eq:startend}), (\ref{eq:datacon}),(\ref{eq:loccon}),(\ref{eq:powercon}),(\ref{eq:taucon}), \nonumber \\
   & (\ref{eq:freqcon}),(\ref{eq:edgcon1}),(\ref{eq:edgcon2}),(\ref{eq:grouptrj}),(\ref{eq:groupoff}). \nonumber
\end{IEEEeqnarray}
For the deterministic counterpart above, we can see that it is still rather complicated and non-convex concerning the trajectory and offloading. To solve the problem effectively, we decompose the problem as one tackles the UAV trajectory and scheduling and the other addresses the power and offloading design, given as
\begin{IEEEeqnarray}{cl}
  \IEEEyesnumber\label{eq:trjsch} \IEEEnosubnumber*
  \min_{\substack{\left\{ \hat{\bm{q}}_n \right\}_{n\in\mathcal{N}}, \left\{ \tau_{n,k}\right\}_{k\in\mathcal{K},n\in\mathcal{N}} \\ \left\{ \mu_n^{(\text{trj})}, \nu_n^{(\text{trj})}\right\}_{n\in\mathcal{N}},\left\{ \mu_{n,k}^{(\text{off})}, \nu_{n,k}^{(\text{off})}\right\}_{k\in\mathcal{K},n\in\mathcal{N}} }} \quad & E \\
   \rm{s.t.} \quad & (\ref{eq:startend}), (\ref{eq:taucon}),(\ref{eq:grouptrj}),(\ref{eq:groupoff}), \nonumber
\end{IEEEeqnarray}
and
\begin{IEEEeqnarray}{cl}
  \IEEEyesnumber\label{eq:pwroff} \IEEEnosubnumber*
  \min_{\substack{\left\{ f_n^{\text{(edg)}} \right\}_{n\in\mathcal{N}}, \left\{p_{n,k}\right\}_{k\in\mathcal{K},n\in\mathcal{N}}\\ \left\{d_{n,k}^{\text{(loc)}}, d_{n,k}^{\text{(off)}}\right\}_{k\in\mathcal{K},n\in\mathcal{N}}, \left\{ \mu_{n,k}^{(\text{off})}, \nu_{n,k}^{(\text{off})}\right\}_{k\in\mathcal{K},n\in\mathcal{N}} }} \quad & E  \\
   \rm{s.t.} \quad & (\ref{eq:datacon}),(\ref{eq:loccon}),(\ref{eq:powercon}), \nonumber \\
   & (\ref{eq:freqcon}),(\ref{eq:edgcon1}),(\ref{eq:edgcon2}),(\ref{eq:groupoff}), \nonumber
\end{IEEEeqnarray}
respectively. The decomposition can be physically interpreted that the problem in~(\ref{eq:trjsch}) addressing the trajectory and scheduling concerns the operations during the whole flight, while the problem in~(\ref{eq:pwroff}) particularly concerns the transmission and offloading behavior in each time interval. Further, the decomposition also facilitates the mathematical analysis as presented below. The problems in~(\ref{eq:trjsch}) and~(\ref{eq:pwroff}) are then solved separately with the optimization variables of the other regarded as constants, and the alternating optimization between them finally reaches the solution to the original problem.

\subsection{Trajectory and Scheduling Optimization}

Consider the problem in~(\ref{eq:trjsch}) with fixed transmit power and offloading strategy, the objective function is now a linear function with respect to the optimization variables. The non-convexity lies in the uncertainty-related offloading constraints in~(\ref{eq:groupoff}). In particular, the non-convex constraint in~(\ref{eq:groupoffa}) is reformulated as
\begin{equation} \label{eq:conphi}
    \begin{aligned}
    &\mathsf{Tr}\left(\bm{A}_{n,k}^{(\text{off})}\right) - \sqrt{-2\ln\left(\rho_{n,k}^{(\text{off})}\right)}\mu_{n,k}^{(\text{off})} \\
    +& \ln\left(\rho_{n,k}^{(\text{off})}\right)\nu_{n,k}^{(\text{off})} - \left\| \hat{\bm{q}}_n - \hat{\bm{w}}_{k} \right\|^2 + \frac{\beta_0p_{n,k}}{\sigma_0^2}\frac{1}{\phi_{n,k}} \ge 0,
    \end{aligned}
\end{equation}
where $ \left\{ \phi_{n,k} \right\}_{\forall k\in\mathcal{K}, \:\forall n\in\mathcal{N}} $ are the introduced auxiliaries with
\begin{equation} \label{eq:defphi}
    {2^{\frac{d_{n,k}^{\text{(off)}}}{\tau_{n,k}}\frac{N}{TB}-1}} \le {\phi_{n,k}}.
    % \frac{1}{2^{\frac{d_{n,k}^{\text{(off)}}}{\tau_{n,k}}\frac{N}{TB}-1}} \ge \frac{1}{\phi_{n,k}}.
\end{equation}
Then, the inequality in~(\ref{eq:conphi}) is convexified through successive convex approximation with respect to $ \phi_{n,k} $ at $ \phi_{n,k}^{\circ} $ as
\begin{equation} \label{eq:conphisca}
    \begin{aligned}
    &\mathsf{Tr}\left(\bm{A}_{n,k}^{(\text{off})}\right) - \sqrt{-2\ln\left(\rho_{n,k}^{(\text{off})}\right)}\mu_{n,k}^{(\text{off})} \\
    +& \ln\left(\rho_{n,k}^{(\text{off})}\right)\nu_{n,k}^{(\text{off})} - \left\| \hat{\bm{q}}_n - \hat{\bm{w}}_{k} \right\|^2 \\
    +& \frac{\beta_0p_{n,k}}{\sigma_0^2} \left( \frac{2}{\phi_{n,k}^{\circ}} - \left(\frac{1}{\phi_{n,k}^{\circ}}\right)^2  {\phi_{n,k}} \right) \ge 0.
    \end{aligned}
\end{equation}
Also, the inequality in~(\ref{eq:defphi}) can be reorganized as
\begin{equation} \label{eq:defphieqv}
    \log_2\left( 1 + \phi_{n,k} \right) \ge \frac{d_{n,k}^{\text{(off)}}N}{TB}\frac{1}{\tau_{n,k}},
\end{equation}
where the joint convexity with respect to $ \phi_{n,k} $ and $ \tau_{n,k} $ is readily seen. With the reformulation above, the problem in~(\ref{eq:trjsch}) is now in the form of
\begin{IEEEeqnarray}{cl}
  \IEEEyesnumber\label{eq:trjschsca} \IEEEnosubnumber*
  \min_{\substack{\left\{ \hat{\bm{q}}_n \right\}_{n\in\mathcal{N}}, \left\{ \tau_{n,k}\right\}_{k\in\mathcal{K},n\in\mathcal{N}} \\ \left\{ \mu_n^{(\text{trj})}, \nu_n^{(\text{trj})}\right\}_{n\in\mathcal{N}} \\ \left\{ \mu_{n,k}^{(\text{off})}, \nu_{n,k}^{(\text{off})}\right\}_{k\in\mathcal{K},n\in\mathcal{N}} \\ \left\{ \phi_{n,k}\right\}_{k\in\mathcal{K},n\in\mathcal{N}} }} \quad & E \\
   \rm{s.t.} \quad & (\ref{eq:startend}), (\ref{eq:taucon}),(\ref{eq:grouptrj}), \nonumber \\
   & (\ref{eq:groupoffb}),(\ref{eq:groupoffc}),(\ref{eq:conphisca}),(\ref{eq:defphieqv}), \nonumber
\end{IEEEeqnarray}
as an approximation at $ \left\{ \phi_{n,k}^{\circ} \right\}_{\forall k\in\mathcal{K}, \:\forall n\in\mathcal{N}} $. Based on the discussion above, we know that the problem in~(\ref{eq:trjschsca}) is convex and thus can be solved efficiently. Finally, we can exploit the SCA procedure to solve a series of problems in~(\ref{eq:trjschsca}) with an updated sequence of $ \left\{ \phi_{n,k}^{\circ} \right\}_{\forall k\in\mathcal{K}, \:\forall n\in\mathcal{N}} $, where the convergence approaches the solution to the trajectory and scheduling optimization in~(\ref{eq:trjsch}).

\subsection{Power and Offloading Optimization}

With fixed trajectory and scheduling, we consider the problem in~(\ref{eq:pwroff}) for power control and offloading design. We can conveniently verify that the objective function is convex and the constraints are also convex other than those in~(\ref{eq:groupoff}). Similarly, we can reformulate the non-convex terms in~(\ref{eq:groupoffa}) as
\begin{equation} \label{eq:conpsi}
    \begin{aligned}
    &\mathsf{Tr}\left(\bm{A}_{n,k}^{(\text{off})}\right) - \sqrt{-2\ln\left(\rho_{n,k}^{(\text{off})}\right)}\mu_{n,k}^{(\text{off})} \\
    +& \ln\left(\rho_{n,k}^{(\text{off})}\right)\nu_{n,k}^{(\text{off})} - \left\| \hat{\bm{q}}_n - \hat{\bm{w}}_{k} \right\|^2 + \frac{\beta_0}{\sigma_0^2}\frac{\varphi_{n,k}^2}{\psi_{n,k}} \ge 0,
    \end{aligned}
\end{equation}
where $ \left\{ \varphi_{n,k} \right\}_{\forall k\in\mathcal{K}, \:\forall n\in\mathcal{N}} $ and $ \left\{ \psi_{n,k} \right\}_{\forall k\in\mathcal{K}, \:\forall n\in\mathcal{N}} $ are the introduced auxiliaries satisfying
\begin{equation} \label{eq:varphidef}
    p_{n,k} \ge \varphi_{n,k}^2,
\end{equation}
and
\begin{equation} \label{eq:psidef}
    2^{\frac{d_{n,k}^{\text{(off)}}}{\tau_{n,k}}\frac{N}{TB}-1} \le \psi_{n,k}.
\end{equation}
The inequalities in~(\ref{eq:varphidef}) and~(\ref{eq:psidef}) are convex. While for the inequality in~(\ref{eq:conpsi}), the left-hand side incorporates concave functions other than the last term, which is a convex function. Then, the SCA technique is employed to approximate the inequality as
\begin{equation} \label{eq:conpsisca}
    \begin{aligned}
    &\mathsf{Tr}\left(\bm{A}_{n,k}^{(\text{off})}\right) - \sqrt{-2\ln\left(\rho_{n,k}^{(\text{off})}\right)}\mu_{n,k}^{(\text{off})} + \ln\left(\rho_{n,k}^{(\text{off})}\right)\nu_{n,k}^{(\text{off})}\\
     - & \left\| \hat{\bm{q}}_n - \hat{\bm{w}}_{k} \right\|^2 + \frac{\beta_0}{\sigma_0^2} \left( \frac{2\varphi_{n,k}^{\circ}}{\psi_{n,k}^{\circ}} \left( \varphi_{n,k} - \varphi_{n,k}^{\circ} \right) \right. \\
    &\qquad\quad \left. + \left(\frac{\varphi_{n,k}^{\circ}}{\psi_{n,k}^{\circ}}\right)^2 \left( \psi_{n,k} - \psi_{n,k}^{\circ} \right) + \frac{\left(\varphi_{n,k}^{\circ}\right)^2}{\psi_{n,k}^{\circ}} \right) \ge 0,
    \end{aligned}
\end{equation}
at $ \left\{ \varphi_{n,k}^{\circ} \right\}_{\forall k\in\mathcal{K}, \:\forall n\in\mathcal{N}} $ and $ \left\{ \psi_{n,k}^{\circ} \right\}_{\forall k\in\mathcal{K}, \:\forall n\in\mathcal{N}} $ to arrive at a convex constraint. Based on the discussion above, the power and offloading problem in~(\ref{eq:pwroff}) can be recast as
\begin{IEEEeqnarray}{cl} \hspace{-5pt}
  \IEEEyesnumber\label{eq:pwroffsca} \IEEEnosubnumber*
  \min_{\substack{\left\{ f_n^{\text{(edg)}} \right\}_{n\in\mathcal{N}}, \left\{p_{n,k}\right\}_{k\in\mathcal{K},n\in\mathcal{N}}\\ \left\{d_{n,k}^{\text{(loc)}}, d_{n,k}^{\text{(off)}}\right\}_{k\in\mathcal{K},n\in\mathcal{N}} \\ \left\{ \mu_{n,k}^{(\text{off})}, \nu_{n,k}^{(\text{off})}\right\}_{k\in\mathcal{K},n\in\mathcal{N}}, \\ \left\{\varphi_{n,k}, \psi_{n,k}\right\}_{k\in\mathcal{K},n\in\mathcal{N}} }} \quad & E  \\
   \rm{s.t.} \quad & (\ref{eq:datacon}),(\ref{eq:loccon}),(\ref{eq:powercon}),
   (\ref{eq:freqcon}),(\ref{eq:edgcon1}),(\ref{eq:edgcon2}), \nonumber \\
   & (\ref{eq:groupoffb}),(\ref{eq:groupoffc}),(\ref{eq:varphidef}),(\ref{eq:psidef}),(\ref{eq:conpsisca}), \nonumber
\end{IEEEeqnarray}
which acts as a convex counterpart approximated at $ \left\{ \varphi_{n,k}^{\circ} \right\}_{\forall k\in\mathcal{K}, \:\forall n\in\mathcal{N}} $ and $ \left\{ \psi_{n,k}^{\circ} \right\}_{\forall k\in\mathcal{K}, \:\forall n\in\mathcal{N}} $ and thus can be solved efficiently. Then, we can solve the original problem in~(\ref{eq:pwroff}) through the SCA procedure where each iteration tackles the problem in~(\ref{eq:pwroffsca}) with continuously updated approximation points.

\begin{algorithm}[t] \label{alg} %\scriptsize
  \caption{Alternating optimization for energy-minimized edge computing in IIoT with UAV jittering}
  Initialization: randomly select $ \Upsilon_0 $ and $ \Omega_0 $ satisfying the constraints in~(\ref{eq:problem})\;
  Derive the deterministic counterparts of the probabilistic constraints in the form of~(\ref{eq:grouptrj}) and~(\ref{eq:groupoff})\;
  \Repeat{$ \left| E_0 - E\left(\Upsilon^{\star}, \Omega^{\star}\right) \right| < \varepsilon_0 $}
  {
    $ E_0 \gets E\left( \Upsilon_0, \Omega_0 \right) $\;
    Initialize $ \left\{ \phi_{n,k}^{\circ} \right\}_{\forall k\in\mathcal{K}, \:\forall n\in\mathcal{N}} $\;
    \Repeat{$ \left| E_{0,1} - E\left( \Upsilon^{\star}, \left\{ \phi_{n,k}^{\star} \right\}_{\forall k\in\mathcal{K}, \:\forall n\in\mathcal{N}}; \Omega_0 \right)  \right| < \varepsilon_{0,1} $}
    {
      $ E_{0,1} \gets E\left( \Upsilon_0, \left\{ \phi_{n,k}^{\circ} \right\}_{\forall k\in\mathcal{K}, \:\forall n\in\mathcal{N}}; \Omega_0 \right) $\;
      Formulate the convex problem in~(\ref{eq:trjschsca}), and find the solution denoted as $ \Upsilon^{\star} $, $ \left\{ \phi_{n,k}^{\star} \right\}_{\forall k\in\mathcal{K}, \:\forall n\in\mathcal{N}} $\;
      $ \phi_{n,k}^{\circ} \gets \phi_{n,k}^{\star} $, $ \forall k\in\mathcal{K}, \:\forall n\in\mathcal{N} $\;
    }
    $ \Upsilon_0 \gets \Upsilon^{\star} $\;
    Initialize $ \left\{ \varphi_{n,k}^{\circ}, \psi_{n,k}^{\circ}\right\}_{\forall k\in\mathcal{K}, \:\forall n\in\mathcal{N}} $\;
    \Repeat{$ \left| E_{0,2} - E\left( \Omega^{\star}, \left\{ \varphi_{n,k}^{\star}, \psi_{n,k}^{\star} \right\}_{\forall k\in\mathcal{K}, \:\forall n\in\mathcal{N}}; \Upsilon_0 \right) \right| < \varepsilon_{0,2} $}
    {
      $ E_{0,2} \gets E\left( \Omega_0, \left\{ \varphi_{n,k}^{\circ}, \psi_{n,k}^{\circ} \right\}_{\forall k\in\mathcal{K}, \:\forall n\in\mathcal{N}}; \Upsilon_0 \right) $\;
      Formulate the convex problem in~(\ref{eq:pwroffsca}), and find the solution denoted as $ \Omega^{\star} $, $ \left\{ \varphi_{n,k}^{\star} \right\}_{\forall k\in\mathcal{K}, \:\forall n\in\mathcal{N}} $, and $ \left\{ \psi_{n,k}^{\star} \right\}_{\forall k\in\mathcal{K}, \:\forall n\in\mathcal{N}} $\;
      $ \varphi_{n,k}^{\circ} \gets \varphi_{n,k}^{\star} $ and $ \psi_{n,k}^{\circ} \gets \psi_{n,k}^{\star} $, $ \forall k\in\mathcal{K}, \:\forall n\in\mathcal{N} $\;
    }
    $ \Omega_0 \gets \Omega^{\star} $\;
  }
\end{algorithm}

\subsection{Overall Algorithm Design}

Based on the discussions above, we first convert the energy-minimization problem into tractable forms with deterministic constraints. Then, the problem is decomposed into two subproblems within an alternating optimization framework, where each subproblem is tackled with the SCA technique~\cite{sca}. Therefore, the algorithm towards energy minimized edge computing considering the UAV jittering can be summarized as Alg.~\ref{alg}. For notation simplicity, we denote the optimization variable group $ \left\{ \left\{ \hat{\bm{q}}_n \right\}_{n\in\mathcal{N}}, \left\{ \tau_{n,k}\right\}_{k\in\mathcal{K},n\in\mathcal{N}} \right\} $ and $ \left\{ \left\{ f_n^{\text{(edg)}} \right\}_{n\in\mathcal{N}}, \left\{p_{n,k}\right\}_{k\in\mathcal{K},n\in\mathcal{N}}, \left\{d_{n,k}^{\text{(loc)}}, d_{n,k}^{\text{(off)}}\right\}_{k\in\mathcal{K},n\in\mathcal{N}} \right\} $ as $ \Upsilon $ and $ \Omega $, respectively. The thresholds defined by $ \varepsilon_0 $, $ \varepsilon_{0,1} $, and $ \varepsilon_{0,2} $ claim the convergence of the iterative processes.

In Alg.~\ref{alg}, Line 6 to 10 correspond to the SCA procedure for the trajectory and scheduling subproblem, and Line 13 to 17 are for the power and offloading subproblem. Then, the alternating optimization between the two subproblems induces the solution to the original energy minimization problem. Note in the algorithm, each iteration for the SCA procedures is to solve a well-defined and bounded convex problem in the form of~(\ref{eq:trjschsca}) or~(\ref{eq:pwroffsca}). The SCA iterations generate a non-increasing sequence with respect to objective function in terms of the overall energy consumption, and thus the convergence is guaranteed.

The complexity of the proposed algorithm is analyzed as follows. Generally, the algorithm features double-layer loops where the outer layer is for alternating optimization framework, while the inner layer incorporates two SCA procedures. We first analyze the complexity to solve problems in~(\ref{eq:trjschsca}) and~(\ref{eq:pwroffsca}) as the iterative process within the SCA procedures. Generally, solving a convex problem with the interior-point method corresponds to the worst-case complexity of $ \mathcal{O}\left(n^{3.5}\log\left(\frac{1}{\epsilon}\right)\right) $, where $ n $ is the number of optimization variables and $ \epsilon $ is the required precision~\cite{IPM}. Then, the problem in~(\ref{eq:trjschsca}) incorporates $ \left(K+2\right)N-2 $ decision variables and $ \left(3K+2\right)N $ slack variables, the computation complexity is then $ \mathcal{O}\left((NK)^{3.5}\log\left(\frac{1}{\epsilon_1}\right)\right) $, with $ {\epsilon_1} $ being the computation precision. Similarly, the problem solving of~(\ref{eq:pwroffsca}) also has complexity of $ \mathcal{O}\left((NK)^{3.5}\log\left(\frac{1}{\epsilon_2}\right)\right) $ with $ {\epsilon_2} $ being the precision, since it has $ \left(3K+1\right)N $ decision variables and $ 4KN $ slack variables. Further assume $ L_1 $ and $ L_2 $ iterations are required to solve the trajectory and scheduling optimization and power and offloading problem, respectively, and $ L $ iterations are needed for the alternating optimization framework, then the overall computation complexity of the proposed algorithm is given as $ \mathcal{O} \left( L\left( L_1 \left((NK)^{3.5}\log\left(\frac{1}{\epsilon_1}\right)\right) + L_2 \left((NK)^{3.5}\log\left(\frac{1}{\epsilon_2}\right)\right) \right) \right) $.

\section{Simulation Results} \label{sec:5}

\begin{figure*}[t] 
  \centerline{
  \subfigure[UAV trajectory along with speed.]{
    \label{fig:2Speed} %% label for first subfigure
    \includegraphics[width=6.0cm]{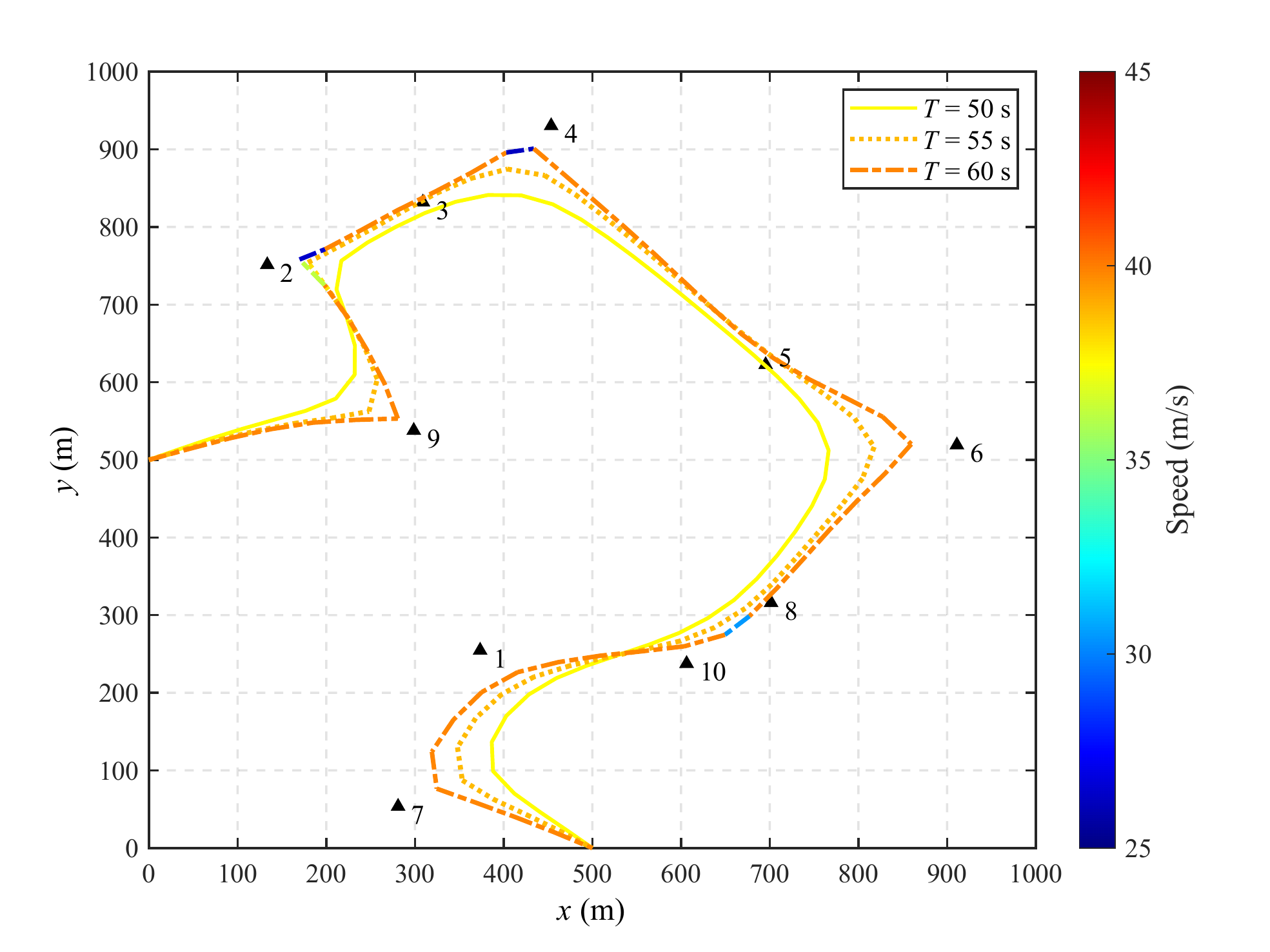}}
  \subfigure[Offloading data amount while UAV flying.]{
    \label{fig:2dOff} %% label for second subfigure
    \includegraphics[width=6.0cm]{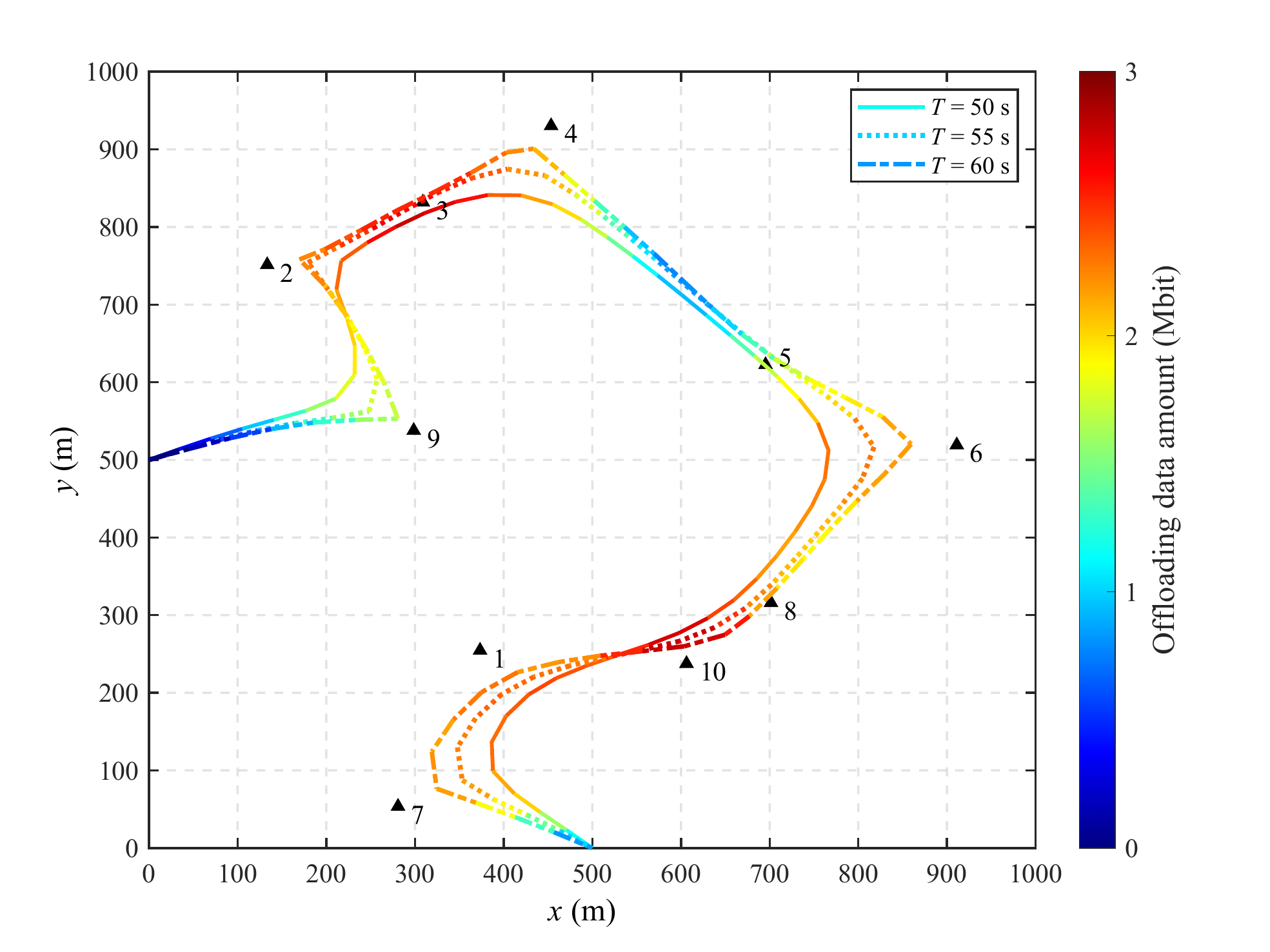}}
    \subfigure[Overall energy consumption and individual data offloading percentage.]{
    \label{fig:2Perfm} %% label for first subfigure
    \includegraphics[width=6.0cm]{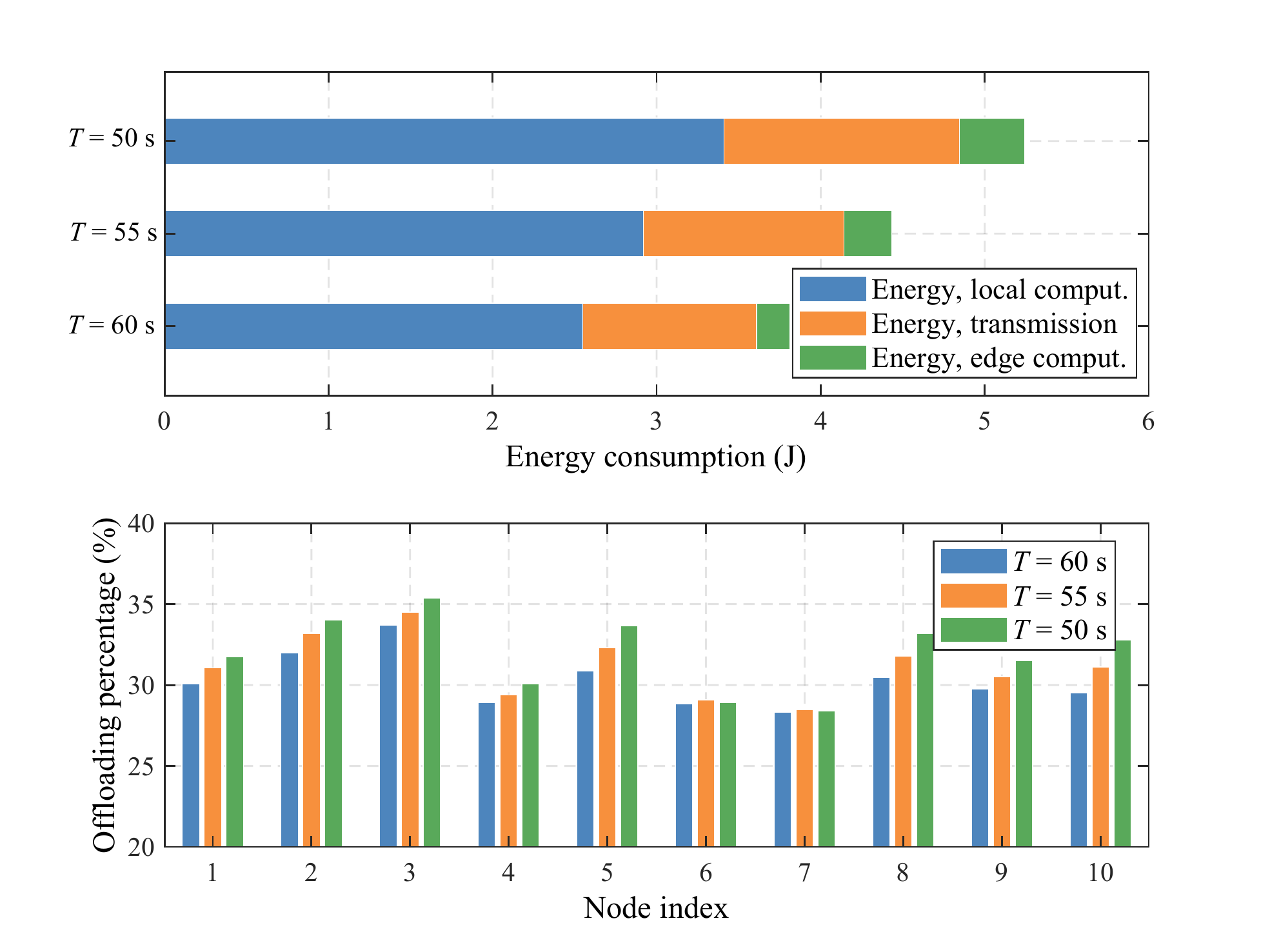}}
  }
  \caption{Performance under different time lengths.}
  \label{fig:perfmT} %% label for entire figure
\end{figure*}

\begin{figure*}[t] %\vspace{-4pt}
  \centerline{
  \subfigure[UAV trajectory along with speed.]{
    \label{fig:3Speed} %% label for first subfigure
    \includegraphics[width=6.0cm]{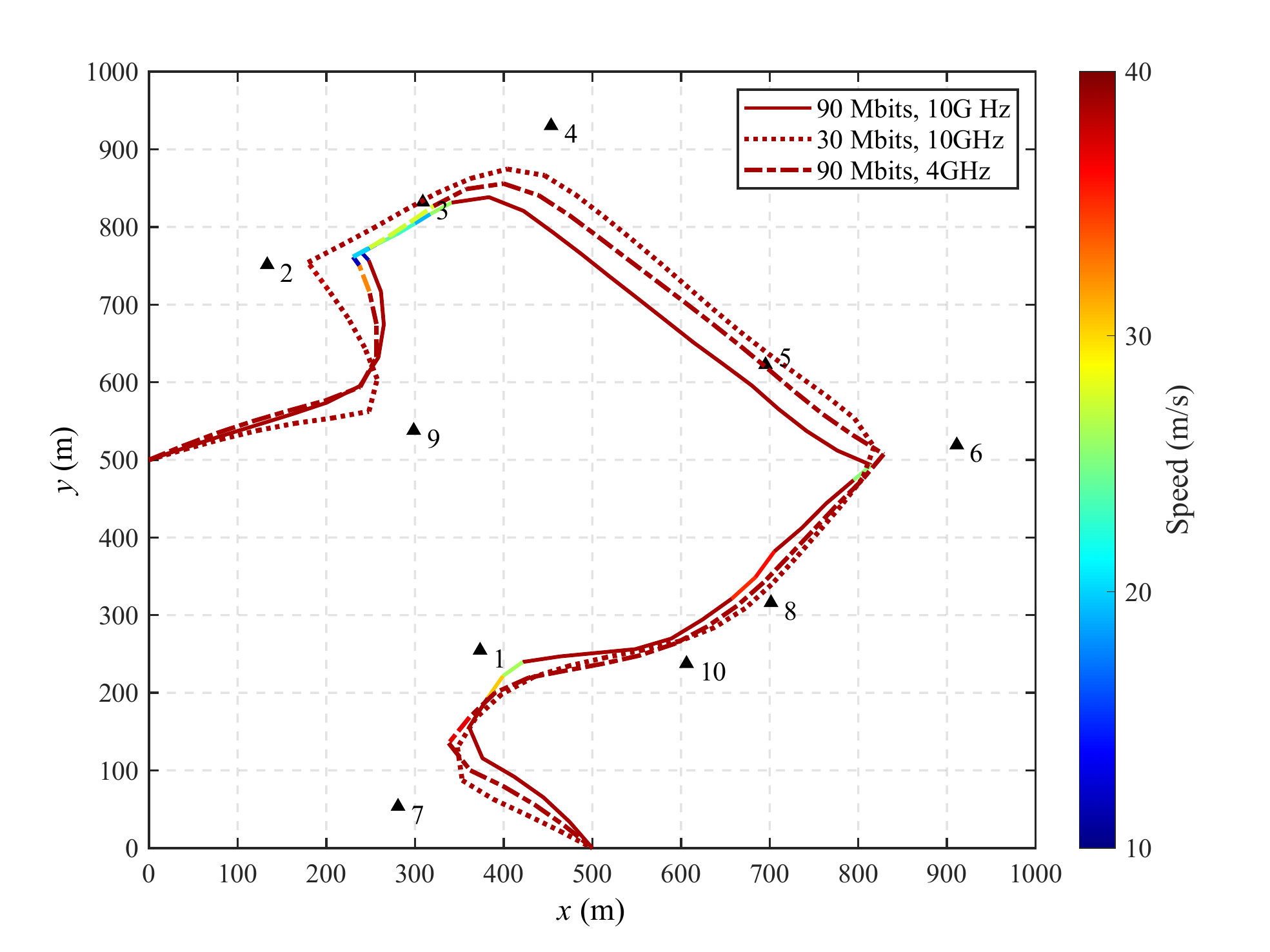}}
  \subfigure[Offloading data amount while UAV flying.]{
    \label{fig:3dOff} %% label for second subfigure
    \includegraphics[width=6.0cm]{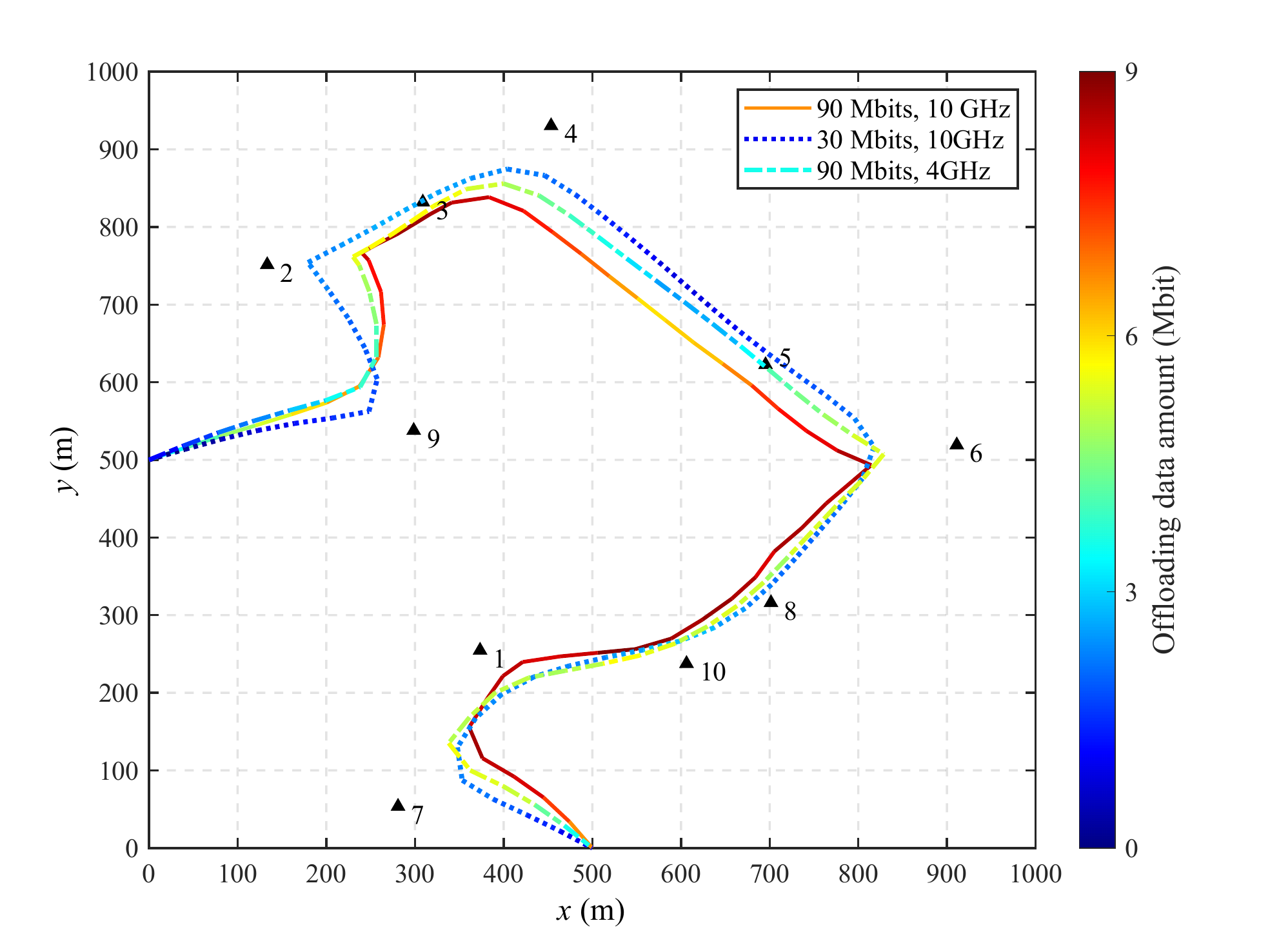}}
    \subfigure[Overall energy consumption and individual data offloading percentage.]{
    \label{fig:3Perfm} %% label for first subfigure
    \includegraphics[width=6.0cm]{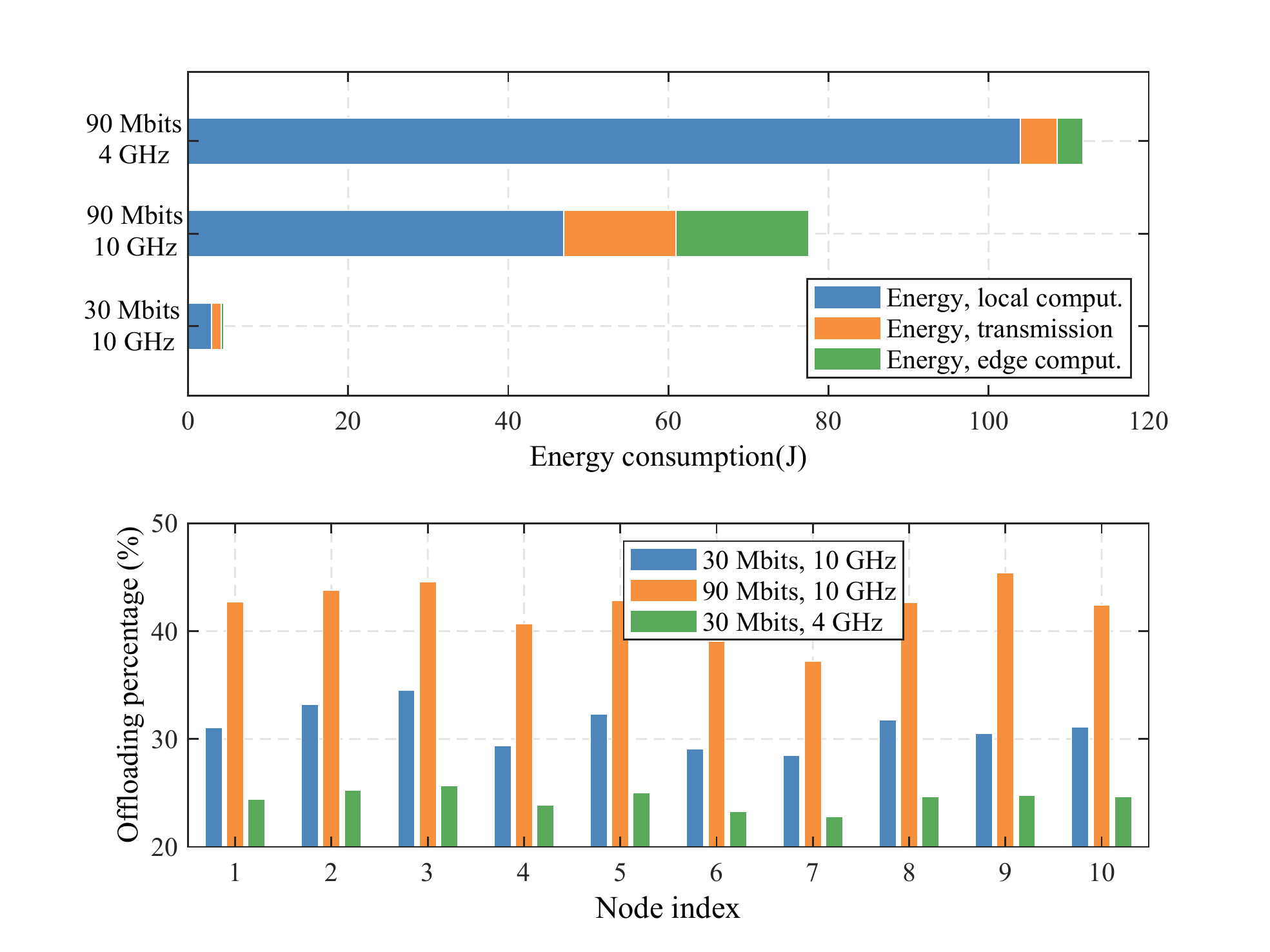}}
  }
  \caption{Performance under different computation data amount and edge computing capability.}
  \label{fig:perfmDC} %% label for entire figure
\end{figure*}

In this section, we provide the simulation results to demonstrate the performance of our proposed scheme. Specifically, we consider an area of 1,000~m $\times$ 1,000~m with randomly located 10 IIoT nodes. As the network topology affects the trajectory design as well as the overall performance, we in the followings use the same topology for fair comparison and clear demonstration. The following parameters are used as defaults unless otherwise noted. We consider a time horizon of 50~s, equally divided into 50~slots. The UAV trajectory has a starting point at (0,~500) and an end point at (500,~0). The UAV is expected to fly at an altitude of 100~m, with the maximum speed of 50~m/s. For the data offloading, the air-ground channel bandwidth is 3~MHz and the attenuation at the reference distance is -60~dB. The maximum transmit power is 800~mW and the noise power is -120~dBm. Meanwhile, each IIoT node has a data amount of 30~Mbits to process and each bit requires 10\textsuperscript{3}~CPU cycles. The highest computation frequency of the node and UAV are 1~GHz and~10 GHz, respectively, and they share the effective capacitance coefficient of~10\textsuperscript{-28}. The weight for edge computing energy consumption is 0.01, and the results below demonstrating the edge computing energy are associated with this coefficient. The UAV waypoints have uncertainty with a standard derivation of 5~m, and the probability of speed violation and offloading data amount violation is~0.1.

Fig.~\ref{fig:perfmT} show the performance of our proposal considering the time length of 50~s, 55~s, and 60~s, during which the UAV finishes the flying as well as the data processing with the IIoT nodes. Specifically, in Figs.~\ref{fig:2Speed} and~{\ref{fig:2dOff}}, we show the UAV trajectory and the offloading data amount during the flying. As can be seen in Fig.~\ref{fig:2Speed}, the UAV trajectory can more closely approach the IIoT nodes with longer flying time, while in contrast, when the trajectory can be more conservative with shorter time length. Also, with longer flying time, the variation of speed during the flight can be more evident, where the locations with denser nodes are of lower speed (node-8,~10; node-2,~3,~4) and locations with sparse nodes (the rest nodes) are of higher speed. Accordingly, we can see in Fig.~\ref{fig:2dOff} that the locations with denser nodes are of higher offloading data amount. For example, when the UAV flies near node-10, it may receive high-volume of offloaded data from node-1, 10, 8, while when near node-5, it may only assist node-5 for the data processing with a smaller size.

In Fig.~\ref{fig:2Perfm}, we show the overall energy consumption and the percentage of offloading data at each individual node. Specifically, in the upper subfigure, we can see that with a tighter time length, the energy consumption becomes higher since the same amount of data needs to be processed in a shorter time. This holds true for both local computing as well as edge computing. In the lower subfigure, we can see that for most nodes, with shorter time lengths, the offloading data percentage becomes higher, as the computation in a shorter time corresponds to a higher requirement for the nodes and thus they seek the UAV edge for assistance. This also responds to the results in the upper subfigure of Fig.~\ref{fig:2Perfm} and Fig.~\ref{fig:2dOff} showing that the shorter time length results in higher edge energy consumption and higher offloading data amount. Moreover, we can see in the lower subfigure of Fig.~\ref{fig:2Perfm}, the nodes closer to the UAV trajectory are of higher offloading percentage while the nodes relatively deviated from the trajectory are of lower offloading ratio. This is as expected since the nodes far from UAV need higher transmission energy for edge computing and thus they are more motivated for local computation.

Fig.~\ref{fig:perfmDC} show the performance with respect to offloading data amount and edge computation capability. Specifically, we in Fig.~\ref{fig:3Speed} show the trajectory along with the flying speed. When the computation requirement is not that urgent, i.e., smaller data amount with a higher edge computing frequency, the trajectory tends to more closely approach the IIoT nodes with more stable flying. In contrast, when the data amount is increased and with lower edge ability, the trajectory becomes more conservative, and the variation in terms of flying speed becomes more evident. Fig.~\ref{fig:3dOff} shows the offloading data amount while flying. As we can see, when the UAV flies near the locations with denser nodes, the offloading data amount becomes higher. Also as expected, with the same edge computing capability, when the overall computation amount is increased, there will be more edge computing data. With the same edge computing capability, the offloading data amount increases with more IIoT data. Revisit the results in~\ref{fig:3Speed}, we can see that though the change in UAV flying speed is not that evident, the difference in offloading data while flying is rather significant.

In Fig.~\ref{fig:3Perfm}, we show the energy consumption and offloading data percentage in the upper and lower subfigures, respectively. Due to the non-linear relationship between energy consumption and computation frequency, the energy requirement is significantly increased with higher data amount. In particular, compared with the case with 30~Mbits node data and 10~GHz edge computing frequency, the local computing, transmission, and edge computing all incur higher energy when the data amount is increased to 90 Mbits. This result is also suggested in Fig.~\ref{fig:3dOff}, as the latter case has a higher offloading data amount. Meanwhile, when the edge computing capability is downgraded to 4~GHz, we can see that the local computing accounts for a large portion of the energy consumption, as the nodes more rely on themselves to process the data rather than resorting to the capability-limited edge. The results in the lower subfigure indicate that the offloading data percentage for each individual node depends on its distance to the trajectory as well as the node density nearby. Meanwhile, we can see interestingly that with the same computation frequency of 10~GHz, when the data amount increased by three times from 30~Mbits to 90~Mbits, the overall offloading percentage increment is only near 10 percent. This indicates that the increment in offloading data is far less than that of computation data, as limited by the computation capability at the edge. Therefore, results in Fig.~\ref{fig:3Perfm} suggest that with a higher edge computation capability to assist data processing, energy consumption can be effectively reduced.

\begin{figure*}[t]  %\vspace{-2pt}
  \centerline{
  \subfigure[UAV trajectory along with speed.]{
    \label{fig:1Speed} %% label for first subfigure
    \includegraphics[width=6.0cm]{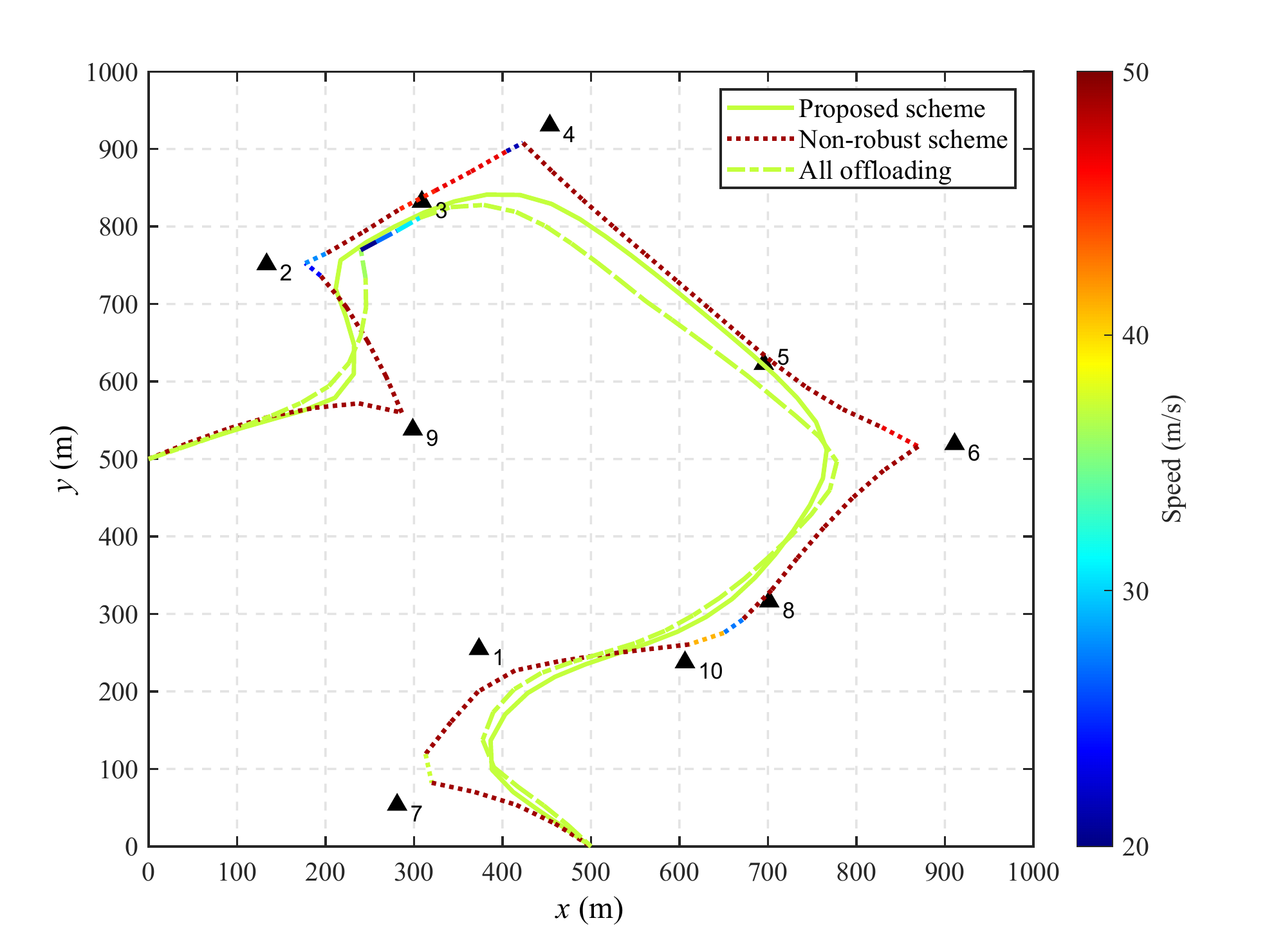}}
  \subfigure[Offloading data amount while UAV flying.]{
    \label{fig:1dOff} %% label for second subfigure
    \includegraphics[width=6.0cm]{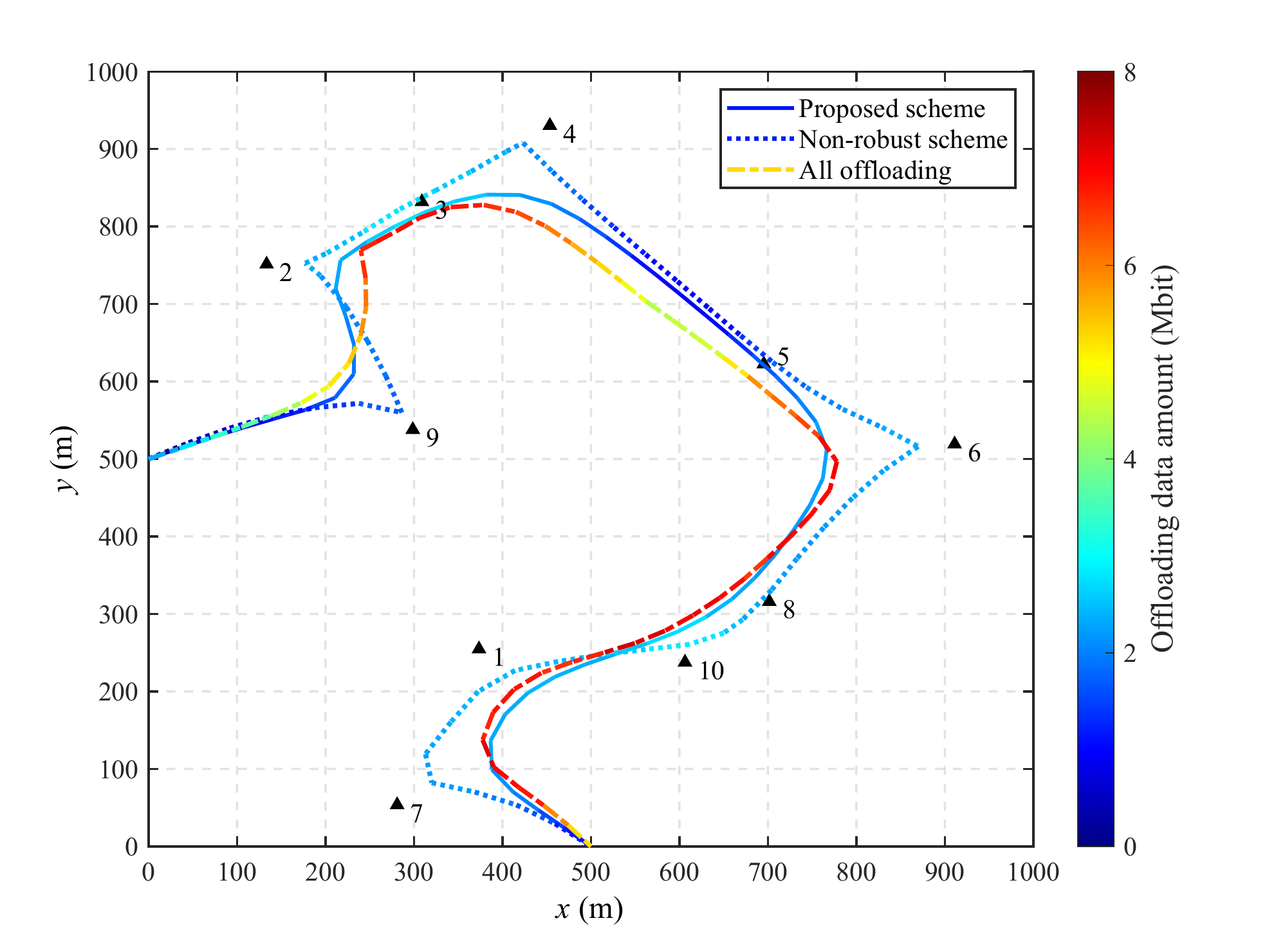}}
    \subfigure[Overall energy consumption with respect to computation data amount.]{
    \label{fig:1Energy} %% label for first subfigure
    \includegraphics[width=6.0cm]{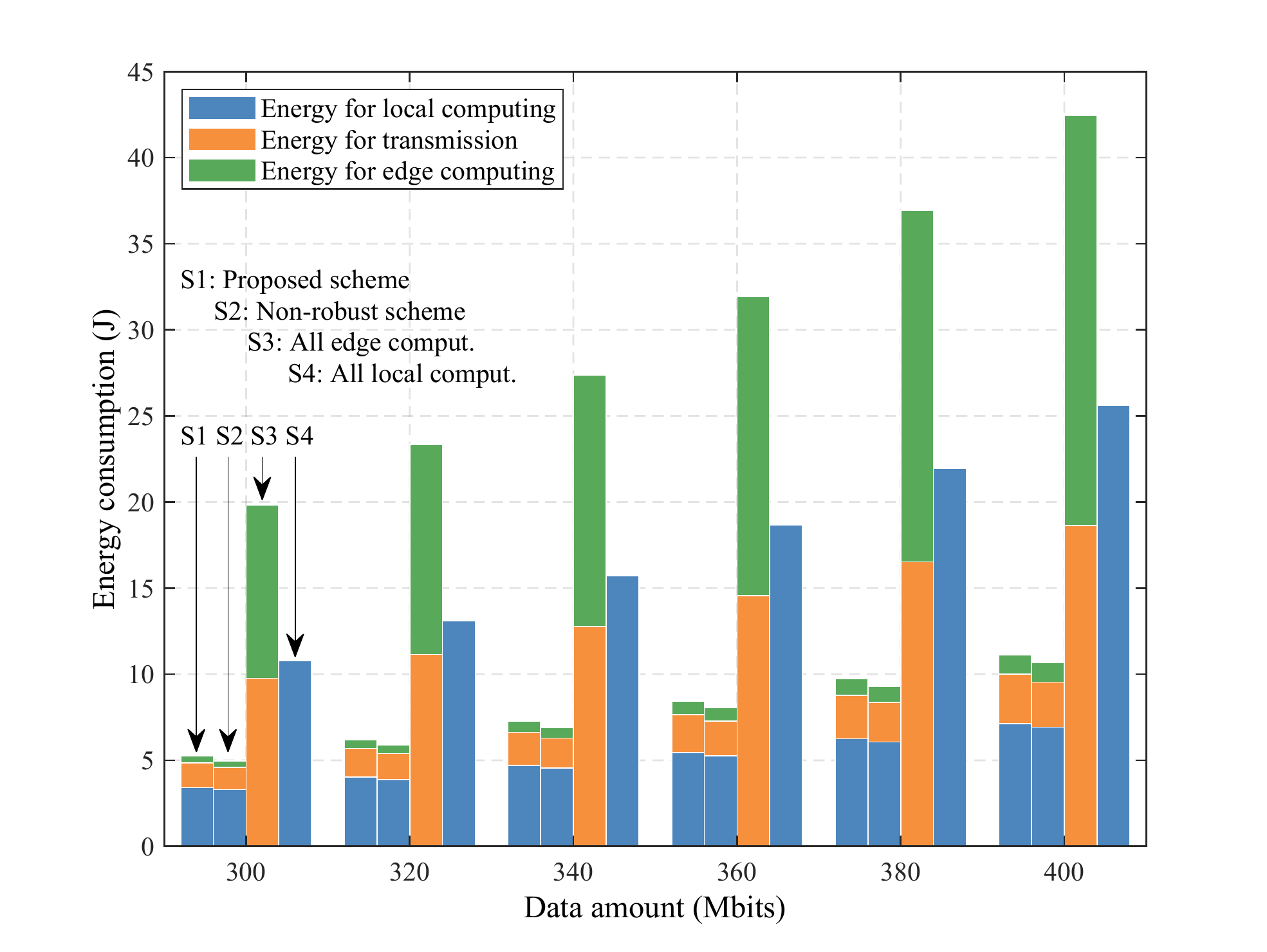}}
  }
    \centerline{
  \subfigure[Histogram of UAV speed with different outage probabilities.]{
    \label{fig:1HistSpeed} %% label for first subfigure
    \includegraphics[width=6.8cm]{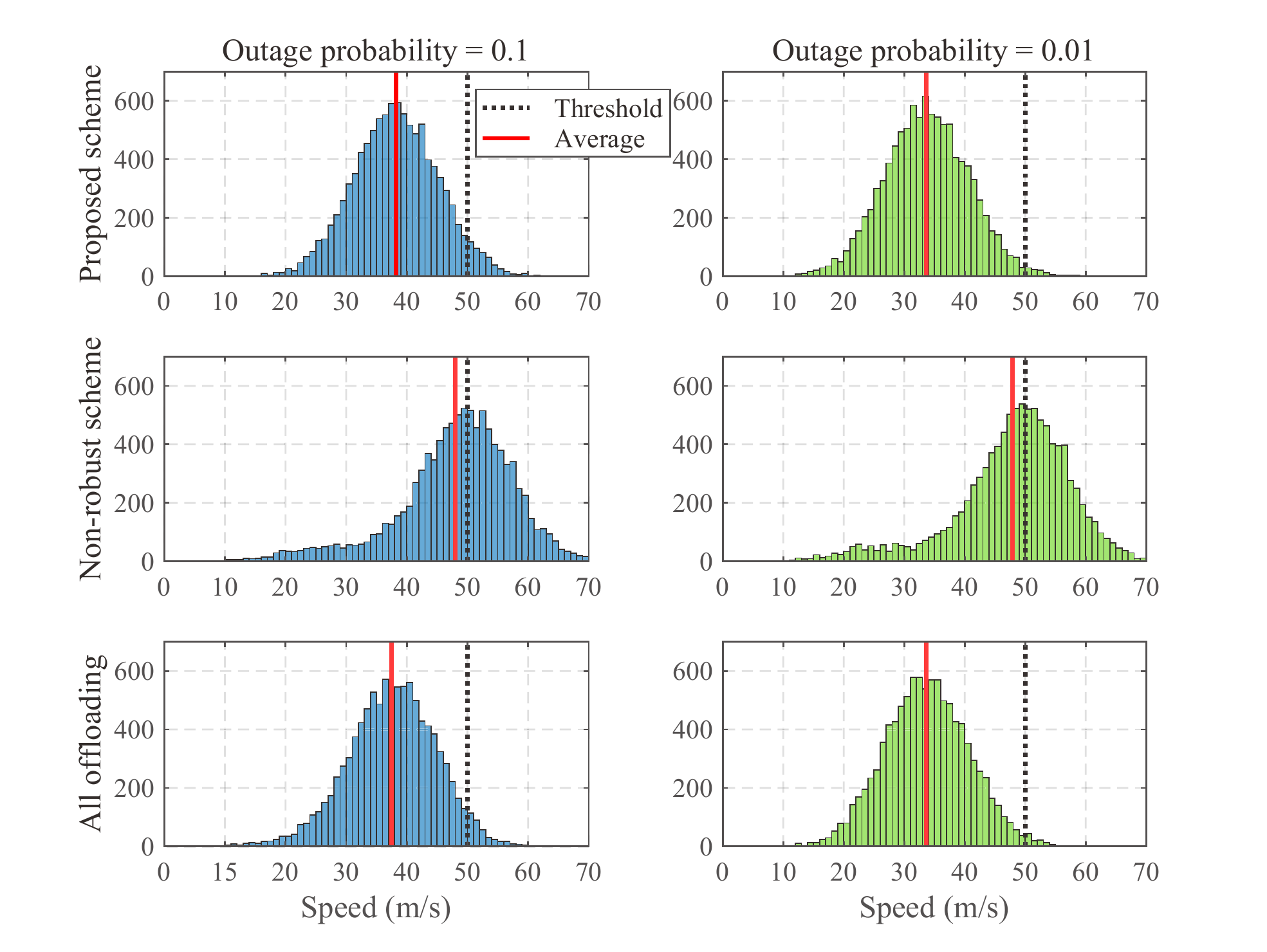}}
  \subfigure[Histogram of offloading completion ratio with different outage probabilities.]{
    \label{fig:1HistdOff} %% label for second subfigure
    \includegraphics[width=6.8cm]{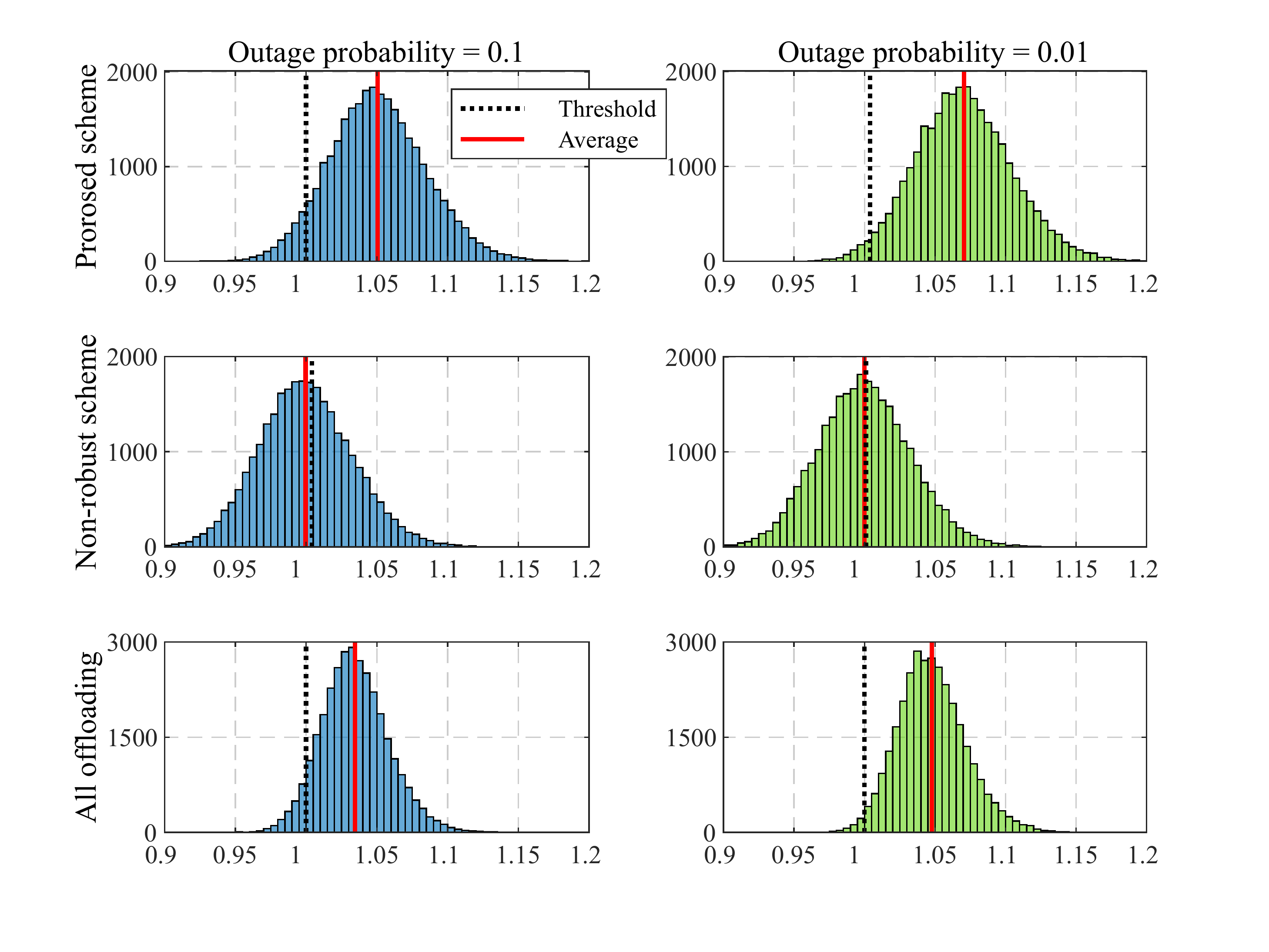}}
  }
  \caption{Performance comparison among different schemes.}
  \label{fig:comp} %% label for entire figure
\end{figure*}

The performance comparison with different schemes is shown in Fig.~\ref{fig:comp}, where we consider our proposal with the case without robust design, i.e., using the deterministic constraints as in~(\ref{eq:speedcon}) and~(\ref{eq:offcon}), rather than~(\ref{eq:speedprob}) and~(\ref{eq:offprob}), to solve the problem, while also evaluating the cases to offload all the data or locally process all the data. In particular, Figs.~\ref{fig:1Speed} and~\ref{fig:1dOff} show the UAV trajectory with the offloading data amount. When there is no consideration on robustness, the UAV generally has a faster flying speed, while the flying with robustness is more conservative, i.e., lower flying speed and less approaching the nodes. Meanwhile, for the case to offload all data, as expected, the flying speed is relatively small while the offloading data amount is rather large. Fig.~\ref{fig:1Energy} shows the energy consumption under different proposals with respect to IIoT data amount. Generally, the energy consumption is higher with more data for all proposals. While compared with the cases of all local or edge computing, where the energy drastically increased with data amount, our proposal that balances the computation induces a much slightly increased energy consumption. Moreover, the overall energy consumption under the non-robust design is slightly smaller than that of our proposal, this is for the reason that due to the UAV jittering, a certain desired amount of data fail to be processed in the system. The amount of unprocessed data under the non-robust design is much larger than that of our proposal, corresponding to lower energy consumption.

In Figs.~\ref{fig:1HistSpeed} and~\ref{fig:1HistdOff}, we show the histogram of the UAV speed and offloading completion ratio under different proposals considering the UAV jittering. The offloading completion ratio is defined as the ratio between the actually transmitted data amount and desired data amount, which should be no less than 1 as a guarantee for data offloading to the edge. In particular, Fig.~\ref{fig:1HistSpeed} shows that the robust designs, i.e., our proposal and all offloading strategy, can effectively guarantee the desired speed violation probability. While in contrast, the non-robust scheme has a much higher chance to violate the maximum speed constraint. Also, the average flying speed under the non-robust scheme is evidently higher than that with robust designs, which is in consistence with the results in Fig.~\ref{fig:1Speed}. Fig.~\ref{fig:1HistdOff} similarly indicates that the non-robust design can easily violate the desired offloading data requirement, whose average completion ratio is also lower than the robust design, while the robust schemes effectively accomplish the data offloading. For our proposal and the case of all offloading, both considering the robustness and thus satisfying the completion subject to the threshold. While in contrast, for the case to offload all the data, the offloading ratio can be more condensed, and the probability with a completion ratio than 1 is high than that of our proposed approach. This indicates that resource required for data offloading is higher than the actual need, thus inducing a more demanding resource consumption.

\section{Conclusion} \label{sec:6}

In this paper, we investigated the UAV-enabled edge computing for IIoT, minimizing the energy consumption with a joint design of trajectory and offloading. We particularly considered the UAV jittering issue and proposed a robust approach to tackle the uncertainties in a probabilistic manner. Simulation results demonstrated that the proposed robust design induced more conservative flying and offloading behavior, and strictly guaranteed the intended outage constraints, whereas the non-robust approach behaved more aggressively while failing the constraints. Also, our proposal effectively reduced energy consumption as compared with the baselines of purely local or edge computing. For future work, we may extend the jittering model to cover more general cases without specifying the distribution information, in order to reach the distributionally robustness in the UAV edge system. Also, the proposed algorithm may be further evaluated in real-world UAV-IIoT scenarios to demonstrate its effectiveness.

\bibliographystyle{IEEEtran}
\bibliography{main}

\end{document}